\documentclass[a4paper,USenglish,cleveref, autoref, thm-restate]{lipics-v2021}

\usepackage{amsmath}
\usepackage{enumitem}
\usepackage{listings}
\usepackage[frozencache]{minted}
\usepackage{plstx}
\usepackage{semantic}
\usepackage{subcaption}
\usepackage{tikz}
\usepackage{wrapfig}
\usepackage{xcolor}

\usepackage[capitalise]{cleveref}

\DeclareMathOperator{\jvp}{jvp}

\DeclareMathOperator{\tuple}{tuple}
\DeclareMathOperator{\vjp}{vjp}

\newcommand*{\djvp}{\overline{\text{\j vp}}}

\makeatletter
\providecommand*{\diff}%
{\@ifnextchar^{\DIfF}{\DIfF^{}}}
\def\DIfF^#1{%
\mathop{\mathrm{\mathstrut d}}%
\nolimits^{#1}\gobblespace}
\def\gobblespace{%
\futurelet\diffarg\opspace}
\def\opspace{%
\let\DiffSpace\!%
\ifx\diffarg(%
\let\DiffSpace\relax
\else
\ifx\diffarg[%
\let\DiffSpace\relax
\else
\ifx\diffarg\{%
\let\DiffSpace\relax
\fi\fi\fi\DiffSpace}

\providecommand*{\deriv}[3][]{%
\frac{\diff^{#1}#2}{\diff #3^{#1}}}

\newcommand*{\Rose}{Rose} % small caps make it look too similar to ROSE
\newcommand*{\code}[1]{\texttt{#1}}
\newcommand*{\codespace}{\ }
\newcommand*{\define}[1]{\textit{#1}}
\newcommand*{\cons}[2]{#1 :: #2}

\newcommand*{\infline}{\\[1ex]}
\newcommand*{\paren}[1]{\code{(}{#1}\code{)}}
\newcommand*{\comma}{\code{,}\codespace}
\newcommand*{\pair}[2]{\paren{{#1}\comma{#2}}}

\newcommand*{\autodifftype}[2]{\mathcal{D}\big[{#1}\big] \rightsquigarrow {#2}}
\newcommand*{\autodiffexpr}[2]{\mathcal{D}\big[{#1}\big] \rightsquigarrow {#2}}
\newcommand*{\autodiffunary}[4]{\mathcal{D}_{#1}\big[{#2}\big] \rightsquigarrow \big\langle {#3} \mathrel{\big|} {#4}\big\rangle}
\newcommand*{\autodiffbinary}[5]{\mathcal{D}_{#1}^{#2}\big[{#3}\big] \rightsquigarrow \big\langle {#4} \mathrel{\big|} {#5}\big\rangle}
\newcommand*{\autodiffblock}[2]{\mathcal{D}\big[{#1}\big] \rightsquigarrow {#2}}

\newcommand*{\transposexpr}[4]{\mathcal{T}_{#1}\big[{#2}\big] \rightsquigarrow \big\langle {#3} \mathrel{\big|} {#4} \big\rangle}
\newcommand*{\transposeblock}[6]{\mathcal{T}_{#1}^{#2}\big[{#3} \mathrel{\big|} {#4}\big] \rightsquigarrow \big\langle {#5} \mathrel{\big|} {#6} \big\rangle}

\newenvironment{codeblock}{
    \begin{center}
    \begin{minipage}{0.85\linewidth}
}{
    \end{minipage}
    \end{center}
}

\newcommand*{\Type}{\code{Type}}
\newcommand*{\Value}{\code{Value}}
\newcommand*{\Index}{\code{Index}}

\newcommand*{\Unit}{\code{()}}
\newcommand*{\Bool}{\code{Bool}}
\newcommand*{\Real}{\code{Real}}
\newcommand*{\Acc}[1]{\code{\&}{#1}}
\newcommand*{\Arr}[2]{\code{[}{#1}\code{]}{#2}}
\newcommand*{\Prod}[2]{\pair{#1}{#2}}

\newcommand*{\Dual}{\code{Dual}}

\newcommand*{\abs}{\code{abs}}
\newcommand*{\sgn}{\code{sgn}}
\newcommand*{\ceil}{\code{ceil}}
\newcommand*{\floor}{\code{floor}}
\newcommand*{\trunc}{\code{trunc}}

\newcommand*{\xnor}{\code{iff}}
\newcommand*{\xor}{\code{xor}}

\newcommand*{\unit}{\code{()}}
\newcommand*{\true}{\code{true}}
\newcommand*{\false}{\code{false}}
\newcommand*{\arr}[1]{\code{[}{#1}\code{]}}
\newcommand*{\tern}[3]{{#1}\codespace\code{?}\codespace{#2}\codespace\code{:}\codespace{#3}}
\newcommand*{\acc}[2]{{#1}\codespace\code{+=}\codespace{#2}}
\newcommand*{\elem}[2]{{#1}\code{[}{#2}\code{]}}
\newcommand*{\fst}[1]{\code{fst}\codespace{#1}}
\newcommand*{\snd}[1]{\code{snd}\codespace{#1}}

\newcommand*{\slice}[2]{\code{\&}{#1}\code{[}{#2}\code{]}}
\newcommand*{\rfst}[1]{\code{\&fst}\codespace{#1}}
\newcommand*{\rsnd}[1]{\code{\&snd}\codespace{#1}}
\newcommand*{\call}[3]{{#1}\code{<}{#2}\code{>(}{#3}\code{)}}
\newcommand*{\map}[3]{\code{[for}\codespace{#1}\code{:}\codespace{#2}\code{,}\codespace{#3}\code{]}}
\newcommand*{\accum}[3]{\code{accum}\codespace{#1}\codespace\code{from}\codespace{#2}\codespace\code{in}\codespace{#3}}

\newcommand*{\bind}[4]{\code{let}\codespace{#1}\code{:}\codespace{#2}\codespace\code{=}\codespace{#3}\codespace\code{in}\codespace{#4}}
\newcommand*{\bindinfer}[3]{\code{let}\codespace{#1}\codespace\code{=}\codespace{#2}\codespace\code{in}\codespace{#3}}

\newcommand*{\sequence}[2]{{#1}\codespace\code{;}\codespace{#2}}

\newcommand*{\decl}[2]{{#1}\code{:}\codespace{#2}}
\newcommand*{\fn}[5]{\code{def}\codespace{#1}\code{<}{#2}\code{>(}{#3}\code{):}\codespace{#4}\codespace\code{=}\codespace{#5}}

\newcommand*{\fty}[3]{\code{<}{#1}\code{>(}{#2}\code{)}\codespace\code{->}\codespace{#3}}

\definecolor{rose}{RGB}{161, 20, 65} % #A11441

\usetikzlibrary{calc}
\definecolor{labelColor}{RGB}{199,58,58}
\newcommand{\ui}[1]{\protect\tikz [font=\sffamily, baseline={($ (current bounding box.center) - (0,.3em) $)}] \fill[fill=labelColor] (0,0em) circle (0.6em) node[text=white] {#1};}

\crefname{lstnumber}{line}{Line}
\crefname{enumi}{}{} % for design choices D1, D2, D3

\lstdefinelanguage{TypeScript}{
  keywords={
    as,
    await,
    const,
    else,
    export,
    for,
    from,
    if,
    import,
    let,
    return,
  },
  sensitive=true,
  string=[b]",
  string=[b]`,
}

\lstdefinelanguage{Rose}{
  keywords={
    accum,
    def,
    for,
    from,
    in,
    let,
    match,
    with,
    type,
  },
  sensitive=true,
}

\hideLIPIcs

\title{Rose: Composable Autodiff for the Interactive Web}

\author{Sam Estep}{Software and Societal Systems Department, Carnegie Mellon University, Pittsburgh, PA, USA \and \url{https://samestep.com} }{estep@cmu.edu}{https://orcid.org/0000-0002-7107-7043}{}

\author{Wode Ni}{Software and Societal Systems Department, Carnegie Mellon University, Pittsburgh, PA, USA \and \url{https://www.cs.cmu.edu/~woden/} }{nimo@cmu.edu}{https://orcid.org/0000-0002-5341-4958}{}

\author{Raven Rothkopf}{Barnard College, Columbia University, New York, NY, USA \and \url{https://ravenrothkopf.com/}}{rgr2124@barnard.edu}{https://orcid.org/0000-0002-3926-683X}{}

\author{Joshua Sunshine}{Software and Societal Systems Department, Carnegie Mellon University, Pittsburgh, PA, USA \and \url{https://www.cs.cmu.edu/~jssunshi/}}{sunshine@cs.cmu.edu}{https://orcid.org/0000-0002-9672-5297}{}

\authorrunning{S Estep, W. Ni, R. Rothkopf, and J. Sunshine}

\Copyright{Sam Estep, Wode Ni, Raven Rothkopf, and Joshua Sunshine}



\funding{This material is based upon work supported by the Aqueduct Foundation and by National Science Foundation under Grant Numbers 1910264, 2119007, and 2150217.}

\acknowledgements{Thanks to Adam Paszke for corresponding about JAX and Dex. The Rose icons were created by Aaron Weiss; we use them via the CC BY 4.0 license.}

\nolinenumbers

\EventEditors{Jonathan Aldrich and Guido Salvaneschi}
\EventNoEds{2}
\EventLongTitle{38th European Conference on Object-Oriented Programming (ECOOP 2024)}
\EventShortTitle{ECOOP 2024}
\EventAcronym{ECOOP}
\EventYear{2024}
\EventDate{September 16--20, 2024}
\EventLocation{Vienna, Austria}
\EventLogo{}
\SeriesVolume{313}
\ArticleNo{33}

\begin{document}

\maketitle

\begin{abstract}
Reverse-mode automatic differentiation (autodiff) has been popularized by deep learning, but its ability to compute gradients is also valuable for interactive use cases such as bidirectional computer-aided design, embedded physics simulations, visualizing causal inference, and more. Unfortunately, the web is ill-served by existing autodiff frameworks, which use autodiff strategies that perform poorly on dynamic scalar programs, and pull in heavy dependencies that would result in unacceptable webpage sizes. This work introduces \Rose, a lightweight autodiff framework for the web using a new hybrid approach to reverse-mode autodiff, blending conventional tracing and transformation techniques in a way that uses the host language for metaprogramming while also allowing the programmer to explicitly define reusable functions that comprise a larger differentiable computation. We demonstrate the value of the \Rose{} design by porting two differentiable physics simulations, and evaluate its performance on an optimization-based diagramming application, showing \Rose{} outperforming the state-of-the-art in web-based autodiff by multiple orders of magnitude.
\end{abstract}

\section{Introduction}\label{sec:introduction}

The web provides a platform for interactive experiences with a uniquely low barrier to usage, because the browser obviates the need for software installation by automatically downloading JavaScript code and running it securely on the client. Industry tools like Google Slides~\cite{google2006} and Figma~\cite{figma2016}, as well as experimental tools like Sketch-n-Sketch~\cite{hempel2019} and Penrose~\cite{ye2020}, leverage this platform to enable authoring of visual media. Interactive explainers like Red Blob Games~\cite{patel2013}, Bartosz Ciechanowski's work~\cite{ciechanowski2014}, and Bret Victor's ``Explorable Explanations''~\cite{victor2011} use the web to help people understand complicated ideas in depth, building up a causal mental model by using sliders to manipulate values and immediately see the effects.

Many of these interactions are fairly simple: often the user just drags a slider back and forth, manipulating a parameter in, for instance, a small physical simulation. But there is room for much richer interactions. An early exploration was g9.js~\cite{webster2016}, which lets the user directly drag around visual shapes, and automatically propagates those changes backward to modify the underlying parameters driving the visualization. This idea of \define{bidirectional editing} or \define{bidirectional transformations}~\cite{anjorin2011} is quite powerful. Some more recent work~\cite{cascaval2022} has explored bidirectional editing in computer-aided design (CAD) via \define{automatic differentiation (autodiff)}, a technique for efficiently computing derivatives of numerical functions. Autodiff has become popularized over the past few years by machine learning (ML) frameworks such as TensorFlow~\cite{abadi2016}, PyTorch~\cite{paszke2019}, and  JAX~\cite{frostig2018}.

Autodiff engines built for ML are focused on high throughput for functions composed of a relatively small number of operations on relatively large tensors. They use \define{reverse-mode} autodiff to compute the gradient of a \define{loss function} in an iterative loop, using a numerical optimization algorithm like stochastic gradient descent or Adam~\cite{kingma2017} to update the parameters of the ML model until the loss value is sufficiently reduced. The loss function is usually chosen to be parallelizable on a GPU. These characteristics do not generally apply to other domains, which often involve \define{scalar programs}~\cite{jakob2022} on which overhead between operations would dominate any tensor-level attempts at parallelism.

For scalar programs, program transformation tools are far more appropriate; examples include Tapenade~\cite{hascoet2013} for Fortran and C, Zygote~\cite{innes2019zygote} for Julia~\cite{bezanson2012}, and Enzyme~\cite{moses2020} for LLVM~\cite{lattner2004}. These tools consume and emit code that deals directly with scalars, reducing expressiveness limitations and operation-level overhead at the expense of the parallelism that ML frameworks gain by specializing to tensor operations. They typically leverage heavy modern compiler technology, using various optimization passes on the program after (and sometimes before) differentiation.

But in the interactive web setting, none of these existing points in the design space are appropriate. We are operating in an environment that is
\begin{itemize}
    \item \textbf{dynamic:} the goal is to let the user author content or build up their mental causal model, by (either implicitly through direct manipulation or explicitly through writing code) specifying a differentiable function themselves. The autodiff engine must operate \emph{online}, differentiating functions directly inside of the user's browser.
    \item \textbf{bandwidth-constrained:} because of the no-install model described above, any JavaScript or WebAssembly code used for autodiff must be shipped over the network to the user's browser. Heavyweight components are unacceptable because their bandwidth requirements would exacerbate page load times beyond the user's patience.
    \item \textbf{latency-constrained:} the system must respond to the user's manipulation of the differentiable function definition, at interactive speed. What we care about is not just the performance of the synthesized gradient, but the sum of that latency with the latency to synthesize the gradient in the first place; \emph{quantitative} differences like a slow ``compilation'' step result in \emph{qualitative} differences in the kinds of interaction possible.
\end{itemize}
All existing autodiff tools, including web-focused tools like TensorFlow.js~\cite{smilkov2019}, fall short on at least one of these constraints: they impose large constant factors for scalar programs, or depend on giant codebases that are difficult to package for the web and result in large bundles, or are too slow to use in an interactive setting, or some combination of these.

To address this gap, we present \textbf{\Rose},\footnote{Not to be confused with the ROSE (all caps) compiler infrastructure.~\cite{quinlan2011}} a scalar-focused autodiff engine for the web that achieves fast compilation time and high generated code performance in a small bundle. As we will describe in \cref{sec:design}, \Rose{} is a hybrid autodiff system~\cite{jakob2022} which blends together techniques from tracing and program transformation before emitting WebAssembly~\cite{haas2017}. Unlike prior program transformation approaches that take advantage of heavyweight compiler optimization toolchains, we produce efficient \Rose{} IR before differentiating by using JavaScript as a \define{metaprogramming environment}, somewhat similar to tracing in popular ML frameworks. But unlike prior tracing approaches that expand all operations into one large graph, we reduce generated code size and thus compilation time by allowing the user to explicitly define \define{composable functions} that can be nested and reused. Our primary contributions are as follows:
\begin{itemize}
    \item We establish the importance of, and constraints imposed by, the interactive web setting, and articulate how those constraints translate to system requirements for autodiff in such a setting.
    \item We describe a novel system design that satisfies these requirements using a careful blend of tracing with program transformation.
    \item We present experiments demonstrating how each component of our design is key to achieving the requirements we have laid out.
    \item We publish \Rose, an open-source software package implementing this design for others to consume and build upon.
\end{itemize}

The rest of this paper is structured as follows. In \cref{sec:background} we give relevant general mathematical background information about autodiff; we introduce a running example that we then implement in \cref{sec:usage}, which discusses \Rose{} from a user perspective. Then \cref{sec:design} discusses the novel design of \Rose, focusing on the high level because that is our more interesting contribution, but also describing some low-level details of autodiff for the curious reader. \Cref{sec:evaluation} describes the experiments we conducted with results showing why this design is key to achieving our design goals. Finally we discuss related work in \cref{sec:related}, and conclude with future work in \cref{sec:conclusion}.

\section{Background}\label{sec:background}

To illustrate the basic ideas of reverse-mode autodiff, we'll walk through the classic example of using gradients to perform linear regression via least-squares optimization. Nothing in this section is new. Almost all the content here can be found in standard textbooks for calculus, linear algebra, and convex optimization; all the rest can be found in the research literature on autodiff~\cite{radul2023}.

Suppose we have $n$ measurements $\mathbf{y} \in \mathbb{R}^n$ of a dependent variable, each corresponding to one of $n$ data points $\mathbf{x}_1, \dots, \mathbf{x}_n \in \mathbb{R}^m$. These data can be assembled into a matrix $\mathbf{X} \in \mathbb{R}^{n \times (m + 1)}$ defined by
\[
    \mathbf{X}
    = \begin{bmatrix}
        1 & \mathbf{x}_1^\top \\
        \vdots & \vdots \\
        1 & \mathbf{x}_n^\top
    \end{bmatrix}
    = \begin{bmatrix}
        1 & x_{11} & \cdots & x_{1m} \\
        \vdots & \vdots & \ddots & \vdots \\
        1 & x_{n1} & \cdots & x_{nm}
    \end{bmatrix}.
\]
We would like to predict the dependent variable as a linear function $\hat{\mathbf{y}} = \mathbf{X}\boldsymbol{\beta}$ where the parameters $\boldsymbol{\beta} \in \mathbb{R}^{m + 1}$ are chosen to minimize the sum of squares of the errors $\boldsymbol{\varepsilon} = \mathbf{y} - \hat{\mathbf{y}}$. That is, one would like to find an optimal solution to the optimization problem
\[
    \min_{\boldsymbol{\beta} \in \mathbb{R}^{m + 1}} f(\boldsymbol{\beta})
    \quad\text{where}\quad
    f(\boldsymbol{\beta})
    = \lVert\boldsymbol{\varepsilon}\rVert^2
    = \lVert\mathbf{y} - \hat{\mathbf{y}}\rVert^2
    = \lVert\mathbf{y} - \mathbf{X}\boldsymbol{\beta}\rVert^2.
\]
Applying convex optimization theory here is standard so we won't belabor it, but because this particular $f$ is differentiable, convex, and smooth, there exists a step size $\eta > 0$ such that if we start with any $\boldsymbol{\beta}_0 \in \mathbb{R}^{m + 1}$ and iteratively compute $\boldsymbol{\beta}_{i + 1} = \boldsymbol{\beta}_i - \eta \nabla f(\boldsymbol{\beta}_i)$, then
\[
    f(\boldsymbol{\beta}_{i + 1}) \leq f(\boldsymbol{\beta}_i)
    \;\forall i \in \mathbb{N},
    \quad\text{and}\quad
    \lim_{i \to \infty} f(\boldsymbol{\beta}_i) \leq f(\boldsymbol{\beta})
    \;\forall \boldsymbol{\beta} \in \mathbb{R}^{m + 1}.
\]
This is gradient descent. Crucially, it depends on being able to compute the gradient $\nabla f$.

\subsection{The vector-Jacobian product}\label{sec:vjp}

As briefly mentioned in \cref{sec:introduction}, reverse-mode autodiff is a general method for computing gradients, which takes in an algorithm to compute a function, and returns an algorithm to compute its gradient. Unlike other approaches to compute derivatives, the power of autodiff lies in its \emph{compositionality} and \emph{efficiency}: we naturally express functions by composing together smaller functions. If we possess an algorithm that computes a given function with a given time complexity, reverse-mode autodiff gives us an algorithm to compute its derivative with the same time complexity, in a way that can be directly composed with derivatives for other functions. For example, in least-squares we compose together three functions
\begin{align*}
    \xi : \mathbb{R}^{m + 1} &\to \mathbb{R}^n &
    \varphi : \mathbb{R}^n &\to \mathbb{R}^n &
    \psi : \mathbb{R}^n &\to \mathbb{R} \\
    \xi(\boldsymbol{\beta}) &= \mathbf{X}\boldsymbol{\beta} &
    \varphi(\hat{\mathbf{y}}) &= \mathbf{y} - \hat{\mathbf{y}} &
    \psi(\boldsymbol{\varepsilon}) &= \lVert\boldsymbol{\varepsilon}\rVert^2
\end{align*}
to form $f = \psi \circ \varphi \circ \xi$. Clearly, the gradient itself is insufficient to express $\nabla f$ compositionally, because $\xi$ and $\varphi$ are not scalar-valued and thus do not have gradients. So we must first have a compositional definition for the derivative.

The \emph{Jacobian} of a function $\mathbf{f} : \mathbb{R}^n \to \mathbb{R}^m$ is the matrix-valued function $\mathbf{J}_{\mathbf{f}} : \mathbb{R}^n \to \mathbb{R}^{m \times n}$ defined by
\[
    \mathbf{J}_{\mathbf{f}}(\mathbf{x})
    = \begin{bmatrix}
        \frac{\partial \mathbf{f}}{\partial x_1} & \cdots & \frac{\partial \mathbf{f}}{\partial x_n}
    \end{bmatrix}
    = \begin{bmatrix}
        \frac{\partial f_1}{\partial x_1} & \cdots & \frac{\partial f_1}{\partial x_n} \\
        \vdots & \ddots & \vdots \\
        \frac{\partial f_m}{\partial x_1} & \cdots & \frac{\partial f_m}{\partial x_n}
    \end{bmatrix}.
\]
From this, if we fix $\mathbf{x} \in \mathbb{R}^n$ then we can define a function $\vjp_{\mathbf{f}}^{\mathbf{x}} : \mathbb{R}^{1 \times m} \to \mathbb{R}^{1 \times n}$, called the \emph{vector-Jacobian product (VJP)}, operating on row vectors called \emph{adjoints} by $\vjp_{\mathbf{f}}^{\mathbf{x}}(\ddot{\mathbf{y}}) = \ddot{\mathbf{y}} \mathbf{J}_{\mathbf{f}}(\mathbf{x})$. In the special case of $m = 1$ we can recover the gradient by $\nabla \mathbf{f}(\mathbf{x}) = \vjp_{\mathbf{f}}^{\mathbf{x}}(1)^\top$, but unlike the gradient, this notion of a derivative actually composes. For instance, if we also have $\mathbf{g} : \mathbb{R}^m \to \mathbb{R}^p$, then
\[
    \vjp_{\mathbf{g} \circ \mathbf{f}}^{\mathbf{x}} = \vjp_{\mathbf{f}}^{\mathbf{x}} \circ \vjp_{\mathbf{g}}^{\mathbf{y}}
    \quad\text{where}\quad
    \mathbf{y} = \mathbf{f}(\mathbf{x}).
\]
This is the chain rule for reverse-mode autodiff, so-called because it composes $\vjp_{\mathbf{f}}$ and $\vjp_{\mathbf{g}}$ in the reverse order of how $\mathbf{f}$ and $\mathbf{g}$ themselves were originally composed. As shown here, computing the derivative $\vjp_{\mathbf{g} \circ \mathbf{f}}$ depends on computing the original function $\mathbf{f}$ itself, so in practice the term ``VJP'' is sometimes actually used to refer to the mapping
\begin{align*}
    \mathbb{R}^n &\to \mathbb{R}^m \times (\mathbb{R}^{1 \times m} \to \mathbb{R}^{1 \times n}) \\
    \mathbf{x} &\mapsto (\mathbf{f}(\mathbf{x}), \vjp_{\mathbf{f}}^{\mathbf{x}})
\end{align*}
that returns both the output of the original function---which we call a \emph{primal} value to contrast it with the VJP's adjoints---and the VJP function.

We'll provide a more general set of composable VJPs in \cref{sec:ir,sec:transpose}, but for this example, we can derive the VJPs
\begin{align*}
    \vjp_\xi^{\boldsymbol{\beta}} : \mathbb{R}^{1 \times n} &\to \mathbb{R}^{1 \times (m + 1)} &
    \vjp_\varphi^{\hat{\mathbf{y}}} : \mathbb{R}^{1 \times n} &\to \mathbb{R}^{1 \times n} &
    \vjp_\psi^{\boldsymbol{\varepsilon}} : \mathbb{R} &\to \mathbb{R}^{1 \times n} \\
    \vjp_\xi^{\boldsymbol{\beta}}(\ddot{\hat{\mathbf{y}}}) &= \ddot{\hat{\mathbf{y}}}\mathbf{X} &
    \vjp_\varphi^{\hat{\mathbf{y}}}(\ddot{\boldsymbol{\varepsilon}}) &= -\ddot{\boldsymbol{\varepsilon}} &
    \vjp_\psi^{\boldsymbol{\varepsilon}}(\ddot{\sigma}) &= 2\ddot{\sigma}\boldsymbol{\varepsilon}^\top
\end{align*}
of the functions we decomposed earlier. One key property to notice here is that, given algorithms to compute $\xi$, $\varphi$, and $\psi$, we immediately have algorithms to compute $\vjp_\xi$, $\vjp_\varphi$, and $\vjp_\psi$, respectively, with the same time complexities. For instance, $\xi$ is $O(mn)$ with na\"ive matrix multiplication, as is $\vjp_\xi$. This is not too surprising, since that is also the same time complexity as directly computing and multiplying by $\mathbf{J}_\xi$. But the time complexity for both $\varphi$ and $\psi$ is $O(n)$, as are the formulas given above for $\vjp_\varphi$ and $\vjp_\psi$, in contrast to the $O(n^2)$ cost of na\"ively computing $\mathbf{J}_\varphi$ or $\mathbf{J}_\psi$. This is because those Jacobians are \emph{sparse}; the ability of reverse-mode autodiff to preserve time complexity in the presence of sparse Jacobians is called the \emph{cheap gradient principle}.

In any case, from these simpler VJPs we can easily compose the gradient of $f$ as
\begin{align*}
    \nabla f(\boldsymbol{\beta})
    = \vjp_f^{\boldsymbol{\beta}}(1)^\top
    &= \vjp_\xi^{\boldsymbol{\beta}}(\vjp_\varphi^{\hat{\mathbf{y}}}(\vjp_\psi^{\boldsymbol{\varepsilon}}(1)))^\top
    = (-2\boldsymbol{\varepsilon}^\top\mathbf{X})^\top
    = 2\mathbf{X}^\top(\mathbf{X}\boldsymbol{\beta} - \mathbf{y}) \\
    &\qquad\text{where}\quad \hat{\mathbf{y}} = \varepsilon(\boldsymbol{\beta}) = \mathbf{X}\boldsymbol{\beta}
    \quad\text{and}\quad \boldsymbol{\varepsilon} = \varphi(\hat{\mathbf{y}}) = \mathbf{y} - \hat{\mathbf{y}}.
\end{align*}

\subsection{The Jacobian-vector product}\label{sec:jvp}

We've talked about the VJP used for reverse-mode autodiff, which is the more useful for optimization, but also the more challenging to implement and specify. \Rose{} allows users to specify custom derivatives for reasons described in \cref{sec:custom}, so to reduce user burden, we allow those custom derivatives to be defined using the simpler \emph{Jacobian-vector product (JVP)} instead of the VJP. In \cref{sec:ir} we'll discuss the actual program transformation used to derive the VJP from the JVP~\cite{radul2023}, but here we first lay out the mathematical groundwork.

For $\mathbf{f} : \mathbb{R}^n \to \mathbb{R}^m$, the JVP of $\mathbf{f}$ at $\mathbf{x} \in \mathbb{R}^n$ is a function $\jvp_{\mathbf{f}}^{\mathbf{x}} : \mathbb{R}^n \to \mathbb{R}^m$ that operates on column vectors called \emph{tangents} by $\jvp_{\mathbf{f}}^{\mathbf{x}}(\dot{\mathbf{x}}) = \mathbf{J}_{\mathbf{f}}(\mathbf{x})\dot{\mathbf{x}}$. In the special case of $n = 1$ we can recover the ordinary derivative by $\mathbf{f}'(\mathbf{x}) = \jvp_{\mathbf{f}}^{\mathbf{x}}(1)$. But more generally, given $\mathbf{g} : \mathbb{R}^m \to \mathbb{R}^p$ we also have a chain rule
\[
    \jvp_{\mathbf{g} \circ \mathbf{f}}^{\mathbf{x}} = \jvp_{\mathbf{g}}^{\mathbf{y}} \circ \jvp_{\mathbf{f}}^{\mathbf{x}}
    \quad\text{where}\quad
    \mathbf{y} = \mathbf{f}(\mathbf{x})
\]
that composes $\jvp_{\mathbf{f}}$ and $\jvp_{\mathbf{g}}$ in the same order as $\mathbf{f}$ and $\mathbf{g}$, hence the name ``forward-mode.'' This is much simpler computationally, because while reverse-mode needed to compose together the VJPs themselves until the end when it could call them with the final adjoint value, forward-mode can simply use the mapping
\begin{align*}
    \mathbb{R}^n \times \mathbb{R}^n &\to \mathbb{R}^m \times \mathbb{R}^m \\
    (\mathbf{x}, \dot{\mathbf{x}}) &\mapsto (\mathbf{f}(\mathbf{x}), \jvp_{\mathbf{f}}^{\mathbf{x}}(\dot{\mathbf{x}}))
\end{align*}
to package together the primal and tangent values.

The judgment is somewhat subjective, but we invite the reader to decide for themselves whether the VJPs derived earlier or these JVPs
\begin{align*}
    \jvp_\xi^{\boldsymbol{\beta}} : \mathbb{R}^{m + 1} &\to \mathbb{R}^n &
    \jvp_\varphi^{\hat{\mathbf{y}}} : \mathbb{R}^n &\to \mathbb{R}^n &
    \jvp_\psi^{\boldsymbol{\varepsilon}} : \mathbb{R}^n &\to \mathbb{R} \\
    \jvp_\xi^{\boldsymbol{\beta}}(\dot{\boldsymbol{\beta}}) &= \mathbf{X}\dot{\boldsymbol{\beta}} &
    \jvp_\varphi^{\hat{\mathbf{y}}}(\dot{\hat{\mathbf{y}}}) &= -\dot{\hat{\mathbf{y}}} &
    \jvp_\psi^{\boldsymbol{\varepsilon}}(\dot{\boldsymbol{\varepsilon}}) &= 2\boldsymbol{\varepsilon}^\top\dot{\boldsymbol{\varepsilon}}
\end{align*}
are closer to the original definitions of $\xi$, $\varphi$, and $\psi$.

When packaging together $\mathbf{f}$ with $\jvp_{\mathbf{f}}$, it is convenient to represent the pair $(\mathbf{x}, \dot{\mathbf{x}}) \in \mathbb{R}^n \times \mathbb{R}^n$ as the single vector $\bar{\mathbf{x}} = \mathbf{x} + \dot{\mathbf{x}}\varepsilon \in \mathbb{D}^n$, making use of the infinitesimal element $\varepsilon$ in the \emph{dual numbers} defined by the commutative algebra $\mathbb{D} = \{a + b\varepsilon \mid a, b \in \mathbb{R}\}$ where $\varepsilon^2 = 0$. For dual numbers $\bar{x} = x + \dot{x}\varepsilon \in \mathbb{D}$ and $\bar{y} = y + \dot{y}\varepsilon \in \mathbb{D}$, the arithmetic operations
\begin{align*}
    \bar{x} + \bar{y} &= x + y + (\dot{x} + \dot{y})\varepsilon \\
    \bar{x} - \bar{y} &= x - y + (\dot{x} - \dot{y})\varepsilon \\
    \bar{x}\bar{y} &= xy + (\dot{x}y + x\dot{y})\varepsilon \\
    \frac{\bar{x}}{\bar{y}} &= \frac{x}{y} + \frac{\dot{x}y - x\dot{y}}{y^2}\varepsilon \qquad\text{where}\quad y \neq 0
\end{align*}
correspond directly to the JVPs of the corresponding arithmetic operations on real numbers. This allows us to define what we'll call the \emph{dual JVP}
\begin{align*}
    \djvp_{\mathbf{f}} : \mathbb{D}^n &\to \mathbb{D}^m \\
    \djvp_{\mathbf{f}}(\mathbf{x} + \dot{\mathbf{x}}\varepsilon) &= \mathbf{f}(\mathbf{x}) + \jvp_{\mathbf{f}}^{\mathbf{x}}(\dot{\mathbf{x}})\varepsilon
\end{align*}
which operates on column vectors of dual numbers. In this framing, specifying the JVPs of our three functions
\begin{align*}
    \djvp_\xi : \mathbb{D}^{m + 1} &\to \mathbb{D}^n &
    \djvp_\varphi : \mathbb{D}^n &\to \mathbb{D}^n &
    \djvp_\psi : \mathbb{D}^n &\to \mathbb{D} \\
    \djvp_\xi(\bar{\boldsymbol{\beta}}) &= \mathbf{X}\bar{\boldsymbol{\beta}} &
    \djvp_\varphi(\bar{\hat{\mathbf{y}}}) &= \mathbf{y} - \bar{\hat{\mathbf{y}}} &
    \djvp_\psi(\bar{\boldsymbol{\varepsilon}}) &= \bar{\boldsymbol{\varepsilon}}^\top\bar{\boldsymbol{\varepsilon}}
\end{align*}
becomes almost trivial. So when we refer to the JVP in \Rose, we're always talking about this dual JVP, not the raw JVP which would operate on real numbers.

\subsection{The Hessian}\label{sec:hessian}

While gradient descent is a first-order method that only makes use of the gradient, other optimization techniques such as Newton's method also need the \emph{Hessian}, which turns $f : \mathbb{R}^n \to \mathbb{R}$ into a matrix-valued function $\mathbf{H}_f : \mathbb{R}^n \to \mathbb{R}^{n \times n}$ yielding all the second-order partial derivatives of $f$ at a given point. The JVP and VJP can be used together to define the Hessian, which is actually just the Jacobian of the gradient; that is, $\mathbf{H}_f = \mathbf{J}_{\nabla f}$. We already know how to use the VJP of a function to compute its gradient. To compute the Jacobian, we just need to observe that the $i$\textsuperscript{th} row of the Jacobian is equal to the JVP applied to the $i$\textsuperscript{th} basis element $\mathbf{e}_i$ of $\mathbb{R}^n$:
\[
    \mathbf{H}_f(\mathbf{x}) = \begin{bmatrix}
        \jvp_{\nabla f}^{\mathbf{x}}(\mathbf{e}_1) & \cdots & \jvp_{\nabla f}^{\mathbf{x}}(\mathbf{e}_n)
    \end{bmatrix}
    \quad\text{where}\quad
    \nabla f(\mathbf{x}) = \vjp_f^{\mathbf{x}}(1)^\top
\]

\section{Using \Rose}\label{sec:usage}

Now that we've discussed the mathematical ideas behind \Rose, in this section we'll describe how programmers use \Rose{} to understand how it fits into the context described in \cref{sec:introduction}. Then in \cref{sec:design} we'll describe our design that blends together tracing with program transformation to fit into this context.

\begin{listing}[t]
\begin{codeblock}
\begin{minted}[linenos,escapeinside=||,]{js}
import { Real, Vec, compile, fn, struct, vjp } from "rose";|\label{line:import_meta}|
import { add, mul, sub, sum } from "rose";|\label{line:import_arith}|
const sqr = (x) => mul(x, x);|\label{line:sqr}|
const leastSquares = ({ m, n }) => fn([{|\label{line:lsq_start}|
  x: Vec(n, Vec(m, Real)), y: Vec(n, Real),
  b0: Real, b: Vec(m, Real),
}], Real, ({ x, y, b0, b }) => sum(n, (i) => {|\label{line:lsq_mid}|
  const yHat = add(b0, sum(m, (j) => mul(x[i][j], b[j])));
  return sqr(sub(y[i], yHat));|\label{line:call_sqr}|
}));|\label{line:lsq_end}|
const linearRegression = async ({ x, y, eta }) => {|\label{line:linreg_start}|
  const [n, m] = [y.length, x[0].length];
  const Beta = struct({ b0: Real, b: Vec(m, Real) });|\label{line:beta}|
  const f = leastSquares({ m, n });|\label{line:lsq_call}|
  const g = fn([Beta], Real, ({ b0, b }) => f({ x, y, b0, b }));|\label{line:partial}|
  const h = fn([Beta], Beta, (beta) => vjp(g)(beta).grad(1));|\label{line:vjp}|
  const grad = await compile(h);|\label{line:compile}|
  let b0 = 0; let b = Array(m).fill(0);|\label{line:zeros}|
  for (;;) {
    const beta = grad({ b0, b });
    const bb0 = b0 - eta * beta.b0;
    const bb = b.map((bi, i) => bi - eta * beta.b[i]);
    if (bb0 === b0 && bb.every((bi, i) => bi === b[i])) break;
    b0 = bb0; b = bb;
  }
  return { b0, b };|\label{line:return}|
};|\label{line:linreg_end}|
console.log(await linearRegression({ eta: 1e-4,|\label{line:call_start}|
  x: [[10],[8],[13],[9],[11],[14],[6],[4],[12],[7],[5]],
  y: [8.04,6.95,7.58,8.81,8.33,9.96,7.24,4.26,10.84,4.82,5.68],
}));|\label{line:call_end}|
\end{minted}
\end{codeblock}
\caption{Using \Rose{} to do linear regression on the first dataset in Anscombe's quartet~\cite{anscombe1973}.}
\label{lst:least-squares}
\end{listing}

\Cref{lst:least-squares} shows a comprehensive end-to-end example using \Rose{} to perform gradient descent to solve the linear regression problem laid out in \cref{sec:background}. This entire example is JavaScript code, which makes use of \Rose{} as a library; lines~\ref{line:import_meta}--\ref{line:import_arith} use standard JavaScript \code{import} statements to pull in definitions from the \Rose{} library. Lines~\ref{line:sqr}--\ref{line:lsq_end} use arithmetic primitives from \Rose{} to implement the loss function from \cref{sec:background}:
\[
    f(\boldsymbol{\beta})
    = \lVert\mathbf{y} - \mathbf{X}\boldsymbol{\beta}\rVert^2
    = \sum_{i = 1}^n \bigg(y_i - \beta_0 - \sum_{j = 1}^m x_{ij} \beta_j\bigg)^2
\]
Lines~\ref{line:lsq_start}--\ref{line:lsq_mid} define the type of the \code{leastSquares} function, which takes as input $\mathbf{x}_1, \dots, \mathbf{x}_n \in \mathbb{R}^m$, as well as $\mathbf{y} \in \mathbb{R}^n$ and $\boldsymbol{\beta} \in \mathbb{R}^{m + 1}$, and returns a scalar. Lines~\ref{line:linreg_start}--\ref{line:linreg_end} wrap around that low-level function to provide a high-level method to perform linear regression, which is then used by lines~\ref{line:call_start}--\ref{line:call_end}. More specifically, line~\ref{line:beta} uses \Rose{} to define the type of $\boldsymbol{\beta} \in \mathbb{R}^{m + 1}$, a vector with $m$ elements plus an additional scalar bias. Line~\ref{line:lsq_call} uses the earlier \code{leastSquares} definition to get a \Rose{} function for the specific $m, n \in \mathbb{N}$ needed, and line~\ref{line:partial} partially applies that function using the provided $\mathbf{x}$ and $\mathbf{y}$ values as constants. Line~\ref{line:vjp} uses \Rose's builtin \code{vjp} function to take the gradient of that partially applied loss function. While mathematically it can be useful to distinguish column and row vectors, the \Rose{} library does not, so the VJP directly produces the gradient. Line~\ref{line:compile} compiles that gradient function to WebAssembly, producing a function that can be called with concrete standard JavaScript values instead of abstract \Rose{} values. Finally, lines~\ref{line:zeros}--\ref{line:return} perform gradient descent by calling the compiled \code{grad} function.

As demonstrated by the above examples, \Rose{} works by letting the user define \emph{differentiable functions} of the form \code{fn(paramTypes, returnType, body)}. We'll discuss this more in \cref{sec:design}, but the high-level idea from a user perspective is that normal JavaScript functions like the one defined on line~\ref{line:sqr} of \cref{lst:least-squares} roughly correspond to what one might think of as \emph{macros} that get expanded on demand to produce code, while \Rose{} functions defined using \code{fn} correspond to traditional functions, and must be well-typed. One could define that \code{sqr} ``macro'' as a function instead as
\begin{codeblock}
\begin{minted}{js}
const sqr = fn([Real], Real, (x) => mul(x, x));
\end{minted}
\end{codeblock}
where the difference is that the body of this function would then be traced only \emph{once} immediately when it is defined, as opposed to the \code{sqr} ``macro'' defined in \cref{lst:least-squares} which gets expanded/traced every time it is called (which in this case happens to only be on line~\ref{line:call_sqr}). This ability for users to choose between these two ways to define functions is a key feature in the novel design of \Rose, and we will see in \cref{sec:evaluation} that it is crucial to achieving the design goals for interactive differentiable web applications that we laid out in \cref{sec:introduction}.

\subsection{Opaque functions}\label{sec:opaque}

\begin{wrapfigure}{R}{0.3\textwidth}
\centering
    \includegraphics[width=0.3\textwidth]{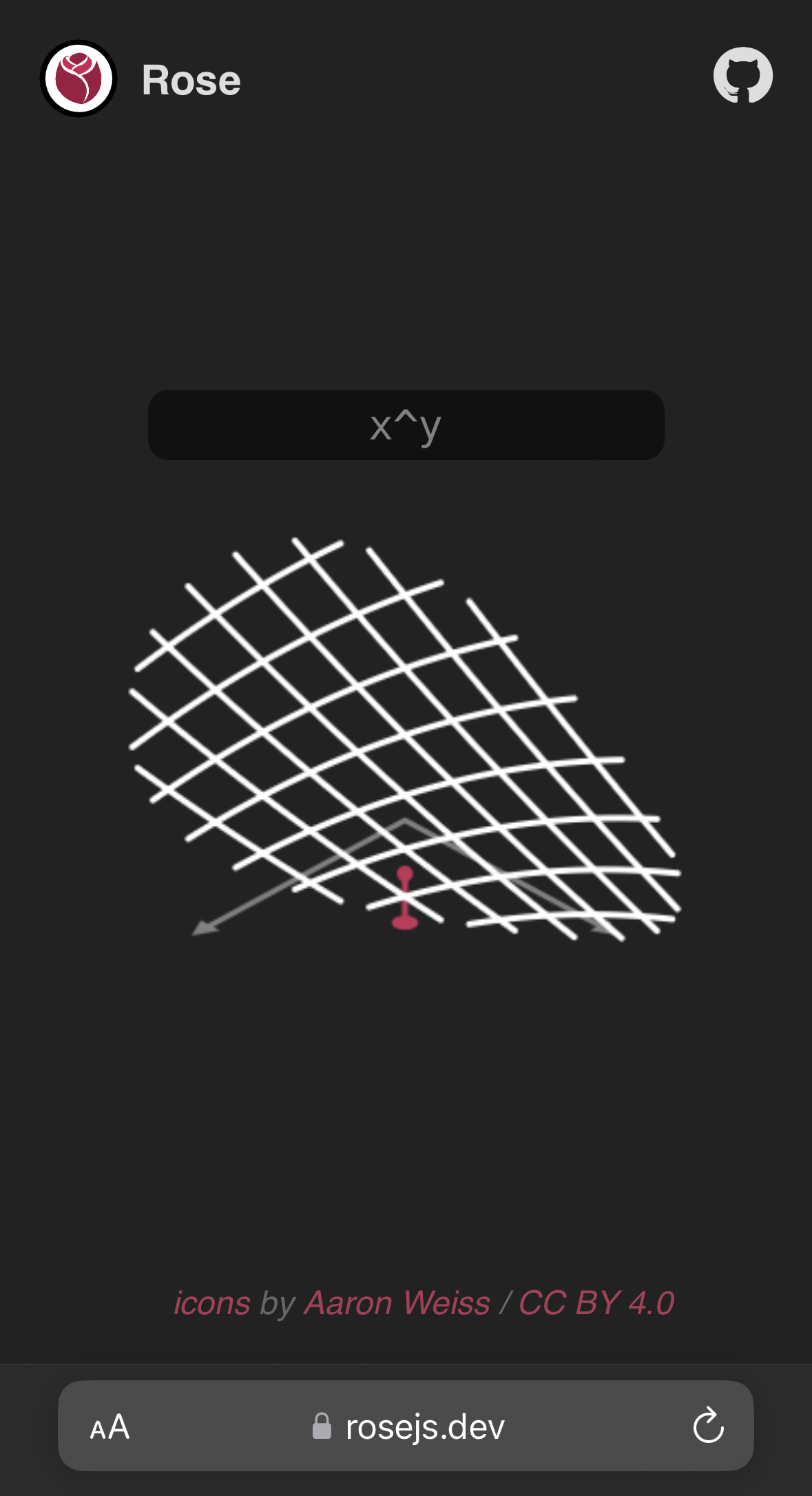}
    \caption{An interactive demo of local quadratic approximation, built with \Rose{} and running in Safari on an iPhone.}
    \label{fig:widget}
\end{wrapfigure}

The above example works well using only builtin arithmetic functions, but it's not interactive; let's look at an interactive example that takes advantage of \Rose's ability to call predefined JavaScript functions and define custom derivatives for them. \Cref{fig:widget} shows an interactive widget on the \Rose{} project website\footnote{\url{https://rosejs.dev/}} displaying the local quadratic approximation to the function $(x, y) \mapsto x^y$, allowing a user to drag the point around to see how the shape of the local quadratic approximation shifts. The page also allows the user to modify the mathematical expression defining the function, causing \Rose{} to immediately re-derive the gradient and Hessian, and compile the new function to WebAssembly. For brevity we omit the code to generate the user interface, and instead focus on how one would use \Rose{} to calculate the first and second derivatives used to visualize the quadratic approximation.

\begin{listing}
\begin{codeblock}
\begin{minted}[linenos,escapeinside=||,]{js}
import { Real, Vec, compile, fn, vjp } from "rose";|\label{line:import_rose}|
import { pow } from "./pow.js";|\label{line:import_pow}|
const R2 = Vec(2, Real);|\label{line:vec2}|
const R22 = Vec(2, R2);|\label{line:mat2}|
const f = fn([R2], Real, ([x, y]) => pow(x, y));|\label{line:f}|
const g = fn([R2], R2, (v) => vjp(f)(v).grad(1));|\label{line:g}|
const h = fn([R2], R22, (v) => {|\label{line:h}|
  const { grad } = vjp(g)(v);|\label{line:grad}|
  return [grad([1, 0]), grad([0, 1])];|\label{line:hess}|
}); |\label{line:h_end}|
const all = fn([Real, Real], { z: Real, g: R2, h: R22 }, (x, y) => {|\label{line:all}|
  const v = [x, y];
  return { z: f(v), g: g(v), h: h(v) };
});|\label{line:all_end}|
const compiled = await compile(all);|\label{line:compile_hess}|
console.log(compiled(2, 3));|\label{line:call}|
\end{minted}
\end{codeblock}
\caption{An example \Rose{} program.}
\label{lst:example}
\end{listing}

\Cref{lst:example} shows a \Rose{} program calculating the value, gradient, and Hessian of the power function at $x = 2$ and $y = 3$. Line~\ref{line:import_pow} imports a power function with a custom derivative, as we'll describe shortly. Lines~\ref{line:vec2}--\ref{line:mat2} define type aliases for $\mathbb{R}^2$ and $\mathbb{R}^{2 \times 2}$, respectively. \Rose{} types are simply JavaScript values, so type aliases are defined using \code{const} in the same way as any other JavaScript value.

Recall that the vector-Jacobian product (VJP) introduced in \cref{sec:vjp} swaps the domain and codomain from the original function. In addition, JavaScript only allows functions to return one argument. Therefore to take the VJP of a \Rose{} function, that function must have only have one parameter. So, line~\ref{line:f} wraps the \code{pow} function to take a single vector argument rather than two scalar arguments, allowing it to be passed to \Rose's \code{vjp} function. Just as we discussed in \cref{sec:vjp}, we compute the gradient by passing in a value of \code{1}.

Lines~\ref{line:h}--\ref{line:h_end} then use the gradient \code{g} of \code{f} to compute its Hessian by differentiating once more. Line~\ref{line:grad} runs the forward pass for the Hessian just once and saves all necessary intermediate values, after which line~\ref{line:hess} runs the backward pass twice with the two basis vectors to compute the full Hessian matrix. Lines~\ref{line:all}--\ref{line:all_end} wrap these three functions into a single function that calls all three and returns the results in a structured form. Finally, line~\ref{line:compile_hess} compiles that function to WebAssembly, and line~\ref{line:call} calls it at the point $(2, 3)$.

\begin{listing}
\begin{codeblock}
\begin{minted}[linenos,escapeinside=||,]{js}
import { Dual, Real, add, div, mul, fn, opaque } from "rose";
const log = opaque([Real], Real, Math.log);|\label{line:log}|
log.jvp = fn([Dual], Dual, ({re:x,du:dx}) => {|\label{line:log_jvp}|
  return { re: log(x), du: div(dx, x) };
});|\label{line:log_jvp_end}|
export const pow = opaque([Real, Real], Real, Math.pow);|\label{line:pow}|
pow.jvp = fn([Dual, Dual], Dual, ({re:x,du:dx}, {re:y,du:dy}) => {
  const z = pow(x, y);
  const dw = add(mul(dx, div(y, x)), mul(dy, log(x)));
  return { re: z, du: mul(dw, z) };
});|\label{line:pow_jvp_end}|
\end{minted}
\end{codeblock}
\caption{The contents of \code{pow.js} defining a differentiable power function.}
\label{lst:opaque}
\end{listing}

\Cref{lst:opaque} shows how the \code{pow} function can be defined to call JavaScript's existing \code{Math.pow} function. Because \Rose{} cannot see the definition of this \code{opaque} function, it must be given a definition for its derivative. Lines~\ref{line:log_jvp}--\ref{line:log_jvp_end} use \Rose{} define the logarithm's dual JVP, which is automatically transposed to produce a VJP as we'll describe in \cref{sec:design}. Specifically, the signature of this function takes the original \code{log} function and maps every instance of the \code{Real} numbers to become the \code{Dual} numbers we introduced in \cref{sec:jvp}. In this case, the returned tangent is given by the familiar rule $\deriv{}{x} \ln x = \frac{1}{x}$ from calculus.

Similarly, lines~\ref{line:pow}--\ref{line:pow_jvp_end} define the power function along with its derivative. Note that, while these two functions use \code{opaque} to define their bodies, they define their derivatives via \code{fn}, the same as the \Rose{} functions we discussed in earlier sections. This means that only the first forward derivative needs to be provided. Since the body of this first derivative is transparent to \Rose, the reverse derivative and any higher derivatives can be computed automatically.

\subsection{Custom derivatives}\label{sec:custom}

\Rose{} lets users define custom derivatives for functions that depend on each other, as in \cref{lst:trig}. The user can also define their own functions to use with \code{opaque}. For instance, one might want to define a \code{print} function for debugging purposes as in \cref{lst:print}, but \Rose{} cannot look inside the definition of \code{print}; by setting \code{print.jvp}, the programmer can tell \Rose{} that the derivative of this function should similarly perform its side effect and otherwise act like the identity function.

\begin{listing}
\begin{codeblock}
\begin{minted}{js}
import { Dual, Real, fn, mul, neg, opaque } from "rose";
const sin = opaque([Real], Real, Math.sin);
const cos = opaque([Real], Real, Math.cos);
sin.jvp = fn([Dual], Dual, ({ re: x, du: dx }) => {
  return { re: sin(x), du: mul(dx, cos(x)) };
});
cos.jvp = fn([Dual], Dual, ({ re: x, du: dx }) => {
  return { re: cos(x), du: mul(dx, neg(sin(x))) };
});
\end{minted}
\end{codeblock}
\caption{Definitions of sine and cosine functions with custom derivatives.}
\label{lst:trig}
\end{listing}

\begin{listing}
\begin{codeblock}
\begin{minted}{js}
import { Dual, Real, fn, opaque } from "rose";
const print = opaque([Real], Real, (x) => {
  console.log(x);
  return x;
});
print.jvp = fn([Dual], Dual, (z) => {
  print(z.re);
  return z;
});
\end{minted}
\end{codeblock}
\caption{A custom \Rose{} function for print debugging.}
\label{lst:print}
\end{listing}

\begin{listing}
\begin{codeblock}
\begin{minted}{js}
import * as rose from "rose";
import { Dual, Real, div, fn, gt, mul, select } from "rose";

const max = (x: Real, y: Real) =>
  select(gt(x, y), Real, x, y);

const sqrt = fn([Real], Real, (x) => rose.sqrt(x));
sqrt.jvp = fn([Dual], Dual, ({ re: x, du: dx }) => {
  const y = sqrt(x);
  const dy = mul(dx, div(1 / 2, max(1e-5, y)));
  return { re: y, du: dy };
});
\end{minted}
\end{codeblock}
\caption{A custom derivative of the square root function to avoid exploding gradients.}
\label{lst:sqrt}
\end{listing}

The other situation is when \Rose{} has automatically constructed a derivative for a function, but that derivative is unstable or otherwise exhibits some undesirable property. \Rose{} allows the user to set a custom derivative for any function, not just \code{opaque} ones. For instance, by default the derivative of the square root function tends to infinity as the argument approaches zero, which causes problems if it is ever called with a zero argument. To prevent this exploding-gradient problem, we sometimes use a square root with a clamped derivative, as in \cref{lst:sqrt}.

In all of these examples, notice that the user only needs to specify the JVP, and not the VJP; this is true even if they later decide to use any of these functions in a VJP context, because \Rose{} uses transposition (described in \cref{sec:transpose}) to automatically construct a VJP from the JVP. A large part of the value of autodiff is that it ensures that the derivative remains in sync with the primal function by construction. Similarly, if we can also assist in keeping the forward-mode and reverse-mode derivatives in sync when one of them must be manually specified, this is a significant benefit for user ergonomics and maintainability.

\section{Design}\label{sec:design}

\begin{figure}
    \includegraphics[width=\textwidth]{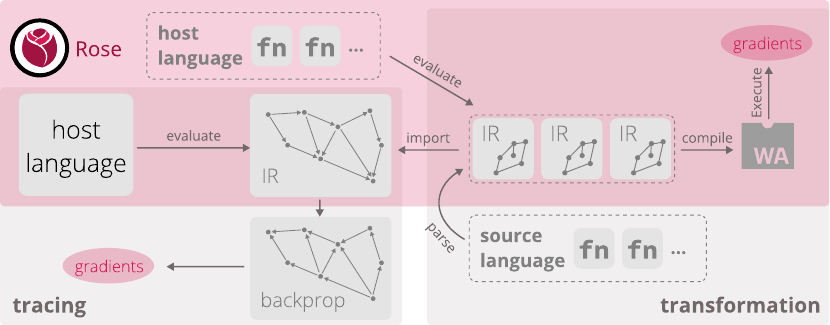}
    \caption{With \textcolor{rose}{\textbf{\Rose}}, the programmer uses the host language for \textit{metaprogramming} like in \textbf{tracing} autodiff, and defines composable \textit{functions} like in \textbf{transformation} autodiff.}
    \label{fig:architecture}
\end{figure}

The previous section described a user experience that hints at the design of \Rose; this section makes that design explicit. Autodiff frameworks typically fall into the two categories of \emph{tracing} and \emph{transformation}, with some \define{hybrid} frameworks combining aspects from the two extremes~\cite{jakob2022}. \Rose{} chooses a specific point in the space of possible hybrid approaches, as diagrammed in \cref{fig:architecture}. To highlight the novel aspects of this design, we will first briefly describe tracing and transformation autodiff; then we will explain how \Rose{} uses parts of both approaches to provide an autodiff engine for the setting described in \cref{sec:introduction}.

In tracing autodiff, the programmer writes code in what we call the \define{host language} (e.g.\ Python~\cite{abadi2016,paszke2019}). They use an autodiff library to construct differentiable scalars, vectors, matrices, and other tensors. Each such tensor can be thought of as a single node in a large \define{computation graph}. Then the programmer calls functions from that autodiff library which take in differentiable tensors and produce more differentiable tensors. Each such function call creates edges in the computation graph from the arguments to the return value. Eventually, a final differentiable scalar value is produced. The programmer calls a special procedure from the autodiff framework, passing in this final scalar value. The autodiff framework traces backward through the computation graph in reverse topological order, attaching gradient values to every node as it goes. The programmer can then use the autodiff framework to access the gradient value attached to any node as they please.

In transformation autodiff, the programmer writes code in what we will call the \define{source language} (e.g.\ Fortran~\cite{hascoet2013} or Julia~\cite{innes2019ssa}). The compiler frontend parses and typechecks this source language to convert it to an \define{intermediate representation (IR)}. This first step typically preserves most of the structure of what the programmer wrote, modulo source formatting. In particular, function definitions and calls in the source text are typically represented as a call graph in an imperative IR, or as lambda terms in a functional IR. Then, the autodiff framework takes in this IR to compute the function the programmer wrote, and emits transformed IR to compute that function along with its gradient. Crucially, this autodiff transformation preserves the asymptotic size of the original program: if the IR representation of the original program has size $n$, then the size of the transformed program to compute gradients is $O(n)$. Then, the compiler backend converts the IR to an executable binary like normal, which can be run to compute the desired gradients.

\Cref{fig:architecture} shows how \Rose{} combines these two approaches. Like tracing, \Rose{} lets the programmer write in a host language they are familiar with: JavaScript, in this case. And like tracing, the programmer is free to use all the features of the host language to describe the shape of their computation. But unlike tracing, and more like transformation, \Rose{} also allows the programmer to explicitly define multiple functions that can be composed together to form a larger computation. Unlike transformation, the programmer's code does not get directly parsed and typechecked to produce the IR; the IR is instead produced by symbolic evaluation like in tracing. But like transformation, the IR can include control flow and function calls, which get explicitly transformed by autodiff rather than being effectively erased as in tracing. And like transformation, the resulting differentiated IR is compiled to WebAssembly~\cite{haas2017} that can then be repeatedly executed to compute gradients for the same program.

By restricting our IR to not allow recursive functions, we are able to use a compilation strategy similar to destination-passing style~\cite{shaikhha2017} to avoid the cost of sophisticated memory management, increasing performance. This strategy not to deallocate memory while computing gradients is justified by the way that reverse-mode autodiff generally needs to retain intermediate values, as described in \cref{sec:transpose}. Importantly, while \Rose{} IR does not allow recursion, the programmer can freely express recursive computations by using the host language for metaprogramming, as we will later discuss in \cref{sec:metaprogramming,sec:dynamism}.

In the following subsections, we will discuss the \Rose{} IR at a theoretical level and explain the autodiff and transposition~\cite{radul2023} program transformations which we use to compute gradients; then in \cref{sec:metaprogramming} we will step back again to discuss how the programmer interacts with this IR indirectly through \Rose{} as a library. All inference rules can be found in \cref{sec:inference}.

\subsection{\Rose{} intermediate representation}\label{sec:ir}

\begin{figure}[t]
    \begin{plstx}
        *: m, n [\in] \mathbb{Z}_{\geq 0} \\
        *: c [\in] \mathbb{R} \\
        : \kappa ::= \Type | \Value | \Index \\
        : \tau ::= t | \Unit | \Bool | \Real | n | \Acc{\tau} | \Arr{\tau}{\tau} | \Prod{\tau}{\tau} | \fty{\underline{\decl{t}{\kappa}}}{\underline{\tau}}{\tau} \\
        : \ominus ::= \lnot | - | \abs | \sgn | \ceil | \floor | \trunc | \code{sqrt} \\
        : \oplus ::= \land | \lor | \xnor | \xor | \neq | < | \leq | = | > | \geq | + | - | \times | \div \\
        : e ::= \unit | \true | \false | c | n | \arr{\underline{x}} | \pair{x}{x} | \ominus x | x \oplus x | \tern{x}{x}{x} | \acc{x}{x} | \elem{x}{x} | \fst{x} | \snd{x} | \slice{x}{x} | \rfst{x} | \rsnd{x} | \call{f}{\underline{\tau}}{\underline{x}} | \map{x}{\tau}{b} | \accum{x}{x}{b} \\
        : b ::= x | \bind{x}{\tau}{e}{b} \\
        : d ::= \fn{f}{\underline{\decl{t}{\kappa}}}{\underline{\decl{x}{\tau}}}{\tau}{b} \\
    \end{plstx}
    \caption{Abstract syntax for \Rose{} IR.}
    \label{fig:syntax}
\end{figure}

\Cref{fig:syntax} shows the abstract syntax for the \Rose{} IR. It is a first-order functional language with non-mutable array types, and a ``reference'' or ``accumulator'' type constructor~\cite{paszke2021}, written $\Acc{\tau}$. \Cref{fig:typing-expr,fig:typing-blocks} show the typing rules for the \Rose{} IR. These are all fairly standard, except for the rules for type constraints $\kappa$. We will explain these less common features of the language in the context of a specific example. \Cref{sec:transpose} will demonstrate the need for accumulators in more detail, but at a high level, they arise naturally because the backward pass of reverse-mode autodiff essentially runs the program backward in time, so reads become accumulation, and writes become reads. Having a type which restricts mutation in this way makes it easier to ensure correctness without introducing additional data dependencies that hinder compiler optimization.

\begin{listing}[t]
\begin{codeblock}
\begin{lstlisting}[language=Rose,numbers=left]
def sum<n: Index>(v: [n]Real): Real =%\label{line:sig}%
  let z: Real = 0.0 in%\label{line:zero}%
  let t: (Real, [n]()) =%\label{line:let_pair}%
    accum a from z in%\label{line:accum}%
      [for i: n,%\label{line:for}%
        let x: Real = v[i] in
        let u: () = a += x in%\label{line:effect}%
        u
      ]%\label{line:for_end}%
    in%\label{line:accum_end}%
  let y: Real = fst t in%\label{line:destructure}%
  y%\label{line:return_sum}%
\end{lstlisting}
\end{codeblock}
\caption{A \Rose{} IR function to compute the sum of a vector of real numbers.}
\label{lst:sum}
\end{listing}

\Rose{} users write JavaScript, so prior examples have been written in JavaScript. However, in this subsection, and \cref{sec:autodiff,sec:transpose}, we describe the IR and so the examples are written in \Rose{} IR. To make this distinction clear, we will format IR examples differently by putting them inside grey boxes.

\cref{lst:sum} computes the sum of a vector of real numbers. Line~\ref{line:sig} says that the function is generic over the size of the array, where the size is represented as a type \code{n} with the \Index{} constraint. The three type constraints in \Rose{} IR are ordered as a hierarchy $\Index <: \Value <: \Type$. Semantically, \Type{} refers to any type that can be stored in a variable; the only types that don't satisfy \Type{} are function types, since \Rose{} IR is first-order. \Value{} is contrasted with reference types: that is, types $\tau : \Type{}$ satisfy the \Value{} constraint unless they are of the form $\tau = \Acc{\tau'}$. Only \Value{} types can be used in other type constructors, so for example, a reference cannot be stored as an element of an array. Finally, the \Index{} constraint marks types that can be used as the index type of an array; the only \Rose{} IR types that satisfy this constraint are those of the form $n \in \mathbb{N}$, which correspond to the value set $\{0, \dots, n - 1\}$.

Line~\ref{line:zero} defines a local called \code{z} of type \Real{} with the value \code{0.0}. This is used by the \code{accum} block on lines~\ref{line:accum} to~\ref{line:accum_end}. This block serves as the scope for the variable \code{a} of type \Acc{\Real}, because \code{z} is of type \Real. The variable \code{a} is in scope for lines~\ref{line:for} to~\ref{line:for_end} and goes out of scope on line~\ref{line:accum_end}. As mentioned above, this is an \define{accumulator} type: it represents a container holding a value of type \Real, but that value cannot be read, and can only be accumulated. In other words, there is no operation of type $\Acc{\Real} \to \Real$. But, given a value of type \Real, one can use the \code{+=} operation to accumulate into the value contained in \code{a}.

With this accumulator in hand, lines~\ref{line:for} to~\ref{line:for_end} execute. These lines are an \define{array constructor} with index type \code{n}, as shown on line~\ref{line:for}. The value type of this array constructor is the unit type \Unit{} so its resulting array type \Arr{\code{n}}{\Unit} holds no data; its sole importance comes from the side effect it performs on line~\ref{line:effect}. The result is that every element of \code{v} gets accumulated into the value of \code{a}, so after this \code{for} block executes, the inner value of \code{a} is equal to the sum of all the values from \code{v}. Again, though, \code{a} cannot actually be read.

Finally, after the body of the \code{accum} block executes, it returns the inner value from \code{a}, along with the value that was returned from the \code{accum} block body itself, which is of type \Arr{\code{n}}{\Unit} as mentioned before. These two items are packaged together into a tuple \code{t} of type \pair{\Real}{\Arr{\code{n}}{\Unit}}. Because no accumulator type can be part of a tuple, this prevents \code{a} itself from escaping from the \code{accum} body, so it is guaranteed to be inaccessible after the \code{accum} block executes. Thus, every accumulator of type $\Acc{\tau}$ starts as accumulate-only, and then when it goes out of scope, its value is guaranteed to be inaccessible except as a ``decayed'' read-only value of type $\tau$. These semantics may seem unintuitive, but they turn out to perfectly model the mapping from forward-mode to reverse-mode autodiff described in \cref{sec:transpose}.

That function definition was quite verbose, as it strictly adhered to the syntax from \cref{fig:syntax} for clarity. In the remainder of this section, we will allow ourselves syntactic sugar to write expressions in places where the strict syntax requires variable names, with the understanding that these could be desugared by introducing intermediate \code{let} bindings:
\begin{codeblock}
\begin{lstlisting}[language=Rose]
def sum<n: Index>(v: [n]Real): Real =
  fst (accum a from 0.0 in [for i: n, a += v[i]])
\end{lstlisting}
\end{codeblock}
In particular, we use this sugared syntax for some inference rules in \cref{fig:forward-mode-block,fig:transpose-expr,fig:transpose-block}.

\subsection{Forward-mode autodiff}\label{sec:autodiff}

As we showed in \cref{sec:jvp}, the forward-mode JVP can often be easier to specify than the reverse-mode VJP, which is why we allow the user to specify custom derivatives using the JVP as described in \cref{sec:custom}. So, we decompose reverse-mode autodiff into two parts~\cite{radul2023}, first applying forward-mode autodiff as we will describe shortly, and then applying a second transformation which we will describe in \cref{sec:transpose}. \Cref{fig:forward-mode-block,fig:forward-mode-type-expr} list inference rules for forward-mode autodiff of \Rose{} IR. Specifically, these rules can be used to transform a function $f$ into a function $f'$ that computes the \emph{dual JVP} of $f$, where the dual numbers are represented in \Rose{} IR by the tuple type $\Dual = \pair{\Real}{\Real}$.

Consider this function, which assumes that $\code{sin} : \fty{}{\Real}{\Real}$ already exists:
\begin{codeblock}
\begin{lstlisting}[language=Rose]
def f(u: Real): Real =
  let v: Real = sin(u) in
  let w: Real = -v in
  w
\end{lstlisting}
\end{codeblock}
We assume that we are already given a dual JVP for \code{sin}. For instance, if the function $\code{cos} : \fty{}{\Real}{\Real}$ also exists, then $\djvp_{\code{sin}} : \fty{}{\Dual}{\Dual}$ might be:
\begin{codeblock}
\begin{lstlisting}[language=Rose]
def jvp_sin((x, dx): Dual): Dual = (sin(x), dx * cos(x))
\end{lstlisting}
\end{codeblock}
Then, by applying the inference rules we have laid out, we get $\djvp_{\code{f}} : \fty{}{\Dual}{\Dual}$:
\begin{codeblock}
\begin{lstlisting}[language=Rose]
def jvp_f(u: Dual): Dual =
  let v: Dual = jvp_sin(u) in
  let (v_re, v_du) = v in
  let w: Dual = (-v_re, -v_du) in
  w
\end{lstlisting}
\end{codeblock}
All the rules for forward-mode autodiff are straightforward and quite standard, so we will not dwell on them here. Now we move on to the more complicated transformation, which maps from forward-mode autodiff to reverse-mode autodiff.

\subsection{Transposition}\label{sec:transpose}

The name ``transposition'' comes from the fact that the JVP and VJP can be thought of as transposes of each other, in the sense of transposing a matrix. Recall that, for $\mathbf{f} : \mathbb{R}^n \to \mathbb{R}^m$ and $\mathbf{x} \in \mathbb{R}^n$, we have
\begin{align*}
    \jvp_{\mathbf{f}}^{\mathbf{x}} : \mathbb{R}^n &\to \mathbb{R}^m &
    \vjp_{\mathbf{f}}^{\mathbf{x}} : \mathbb{R}^{1 \times m} &\to \mathbb{R}^{1 \times n} \\
    \jvp_{\mathbf{f}}^{\mathbf{x}}(\dot{\mathbf{x}}) &= \mathbf{J}_{\mathbf{f}}(\mathbf{x})\dot{\mathbf{x}} &
    \vjp_{\mathbf{f}}^{\mathbf{x}}(\ddot{\mathbf{y}}) &= \ddot{\mathbf{y}} \mathbf{J}_{\mathbf{f}}(\mathbf{x})
\end{align*}
which both make use of the Jacobian matrix $\mathbf{J}_{\mathbf{f}}(\mathbf{x}) : \mathbb{R}^{m \times n}$. So, $\mathbf{J}_{\mathbf{f}}(\mathbf{x})$ is the matrix of the linear transformation $\jvp_{\mathbf{f}}^{\mathbf{x}}$. But then, observe that $\mathbf{J}_{\mathbf{f}}(\mathbf{x})^\top : \mathbb{R}^{n \times m}$ is the matrix of the linear transformation
\begin{align*}
    \mathbb{R}^m &\to \mathbb{R}^n \\
    \dot{\mathbf{y}}
    &\mapsto \mathbf{J}_{\mathbf{f}}(\mathbf{x})^\top \dot{\mathbf{y}}
    = (\dot{\mathbf{y}}^\top \mathbf{J}_{\mathbf{f}}(\mathbf{x}))^\top
    = \vjp_{\mathbf{f}}^{\mathbf{x}}(\dot{\mathbf{y}}^\top)^\top.
\end{align*}
If we had an explicit dense representation of $\mathbf{J}_{\mathbf{f}}$ then it would be trivial to transpose. But we don't; instead, we have an implicit, potentially sparse, representation of $\mathbf{J}_{\mathbf{f}}$ in the form of the dual JVP function $\djvp_{\mathbf{f}}$. \Cref{fig:transpose-expr,fig:transpose-block} list inference rules to transpose the linear derivative represented by the dual JVP in \Rose{} IR. This transformation is considerably more complicated than the initial transformation to do forward-mode autodiff. We will walk through this transformation using the example from \cref{sec:autodiff}, leaving a more systematic exposition to the prior literature~\cite{paszke2021,radul2023} on which our approach was based.

Note that in our inference rules, we attach semantic meaning to certain special notation on variable names. The key idea in transposition is that we split the dual JVP into two functions, a forward pass and a backward pass, where the backward pass transforms read-only types $\tau$ to accumulate-only types $\Acc{\tau}$, and vice versa. As we described in \cref{sec:ir}, each accumulate-only value of type $\Acc{\tau}$ eventually ``decays'' into a read-only value of type $\tau$. In general we use two dots to denote accumulators. So we have two cases:
\begin{itemize}
    \item For $x : \tau$, the forward pass only has the primal variable $x$; to represent the adjoint of $x$, the backward pass has $\ddot{x} : \Acc{\tau}$, which decays into $\dot{x} : \tau$.
    \item For $\ddot{x} : \Acc{\tau}$, the forward pass has the primal accumulator $\ddot{x}$ which decays into $x : \tau$; to represent the adjoint of $x$, the backward pass only has $\dot{x} : \tau$.
\end{itemize}
So in either case, every variable before transposition always conceptually becomes three variables $x, \dot{x} : \tau$ and $\ddot{x} : \Acc{\tau}$. The value adjoint $\dot{x}$ only ever exists in the backward pass. The accumulator $\ddot{x}$ exists in either the forward pass or the backward pass depending on whether the original variable was an accumulator or not, but never in both passes. And the primal $x$ generally exists in both passes, because the backward pass often needs to refer to primal values to perform calculations; this is the case with multiplication and division, for example.

The other notation is the usage of a hat on a variable like $\hat{x} : \Real$ to denote that it exclusively represents a tangent or adjoint value, rather than a primal value or a combination of both using dual numbers. This is the same notation used in \cref{fig:forward-mode-block}.

\begin{listing}[t]
\begin{codeblock}
\begin{lstlisting}[language=Rose]
type Tf = (Dual,(Dual,(Tape_sin,(Real,(Real,(Dual,()))))))
def fwd_f(u: Dual): (Dual, Tf) =
  let (v, t0) = fwd_sin(u) in
  let (v_re, v_du) = v in
  let w_re = -v_re in
  let w_du = 0 in
  let w = (w_re, v_re) in
  (w, (v, (t0, (w_re, (w_du, (w, ()))))))
def bwd_f(ddu: &Dual, dw0: Dual, t: Tf): () =
  let (v, (t0, t1)) = t in
  let (dv, ()) = accum ddv from v in (
    let v_re = fst v in
    let ddv_re = &fst ddv in
    let v_du = snd v in
    let ddv_du = &snd ddv in
    let (w_re, t2) = t1 in
    let (dw_re, ()) = accum ddw_re from w_re in (
      let (w_du, t3) = t2 in
      let (dw_du, ()) = accum ddw_du from w_du in (
        let (w, t4) = t3 in
        let (dw1, ()) = accum ddw from w in (
          ddw += dw0
        ) in
        let (dw1_re, dw1_du) = dw1 in
        ddw_re += dw1_re ;
        ddw_du += dw1_du
      ) in
      ddv_du += -dw_du
    ) in
    ()
  ) in
  bwd_sin(ddu, dv, t0)
\end{lstlisting}
\end{codeblock}
\caption{Strict transposition of the function \code{f} from \cref{sec:autodiff}.}
\label{lst:transpose-example}
\end{listing}

Now, back to the example. By strictly following the inference rules we have laid out, we end up with \cref{lst:transpose-example}. As is common with program transformations like this, the resulting code is highly redundant; via a straightforward optimization pass that we omit here for brevity, we obtain the following:
\begin{codeblock}
\begin{lstlisting}[language=Rose]
def fwd_f((u, _): Dual): (Dual, Tape_sin) =
  let ((v, _), t) = fwd_sin((u, 0)) in
  let w = -v in
  ((w, 0), t)
def bwd_f(ddu: &Dual, (_, dw): Dual, t: Tape_sin): () =
  let dv = -dw in
  bwd_sin(ddu, (0, dv), t)
\end{lstlisting}
\end{codeblock}
This example hints at the fact that the infinitesimal part is always zero for dual numbers representing primal values, and the real part is always zero for dual numbers representing adjoint values. Thus, in our actual system, in the aforementioned optimization pass we also translate the \Dual{} type back to \Real{}, essentially reversing our replacement of \Real{} with \Dual{} from \cref{sec:autodiff}:
\begin{codeblock}
\begin{lstlisting}[language=Rose]
def fwd_f(u: Real): (Real, Tape_sin) =
  let (v, t) = fwd_sin(u) in
  let w = -v in
  (w, t)
def bwd_f(ddu: &Real, dw: Real, t: Tape_sin): () =
  let dv = -dw in
  bwd_sin(ddu, dv, t)
\end{lstlisting}
\end{codeblock}
Thus concludes the transposition of \code{f}. Similarly we can also transpose \code{jvp\_sin}:
\begin{codeblock}
\begin{lstlisting}[language=Rose]
def fwd_sin(x: Real): (Real, Real) =
  (sin(x), cos(x))
def bwd_sin(ddx: &Real, dy: Real, z: Tape_sin): () =
  ddx += dy * z
\end{lstlisting}
\end{codeblock}

\subsection{Metaprogramming}\label{sec:metaprogramming}

\begin{figure}
\centering
\begin{subfigure}{0.54\textwidth}
\begin{codeblock}
\begin{minted}{js}
import { Real, add, div, fn } from "rose";
const f = (x) =>
  fn([Real], Real, (y) => div(x, y));
const g = (y) =>
  fn([Real], Real, (x) => div(x, y));
let [f0, f1, f2] = [f(5), f(7), f(5)];
fn([Real], Real, (z) => add(f1(z), f1(z)));
\end{minted}
\end{codeblock}
\end{subfigure}
\hfill
\begin{subfigure}{0.41\textwidth}
\begin{codeblock}
\begin{lstlisting}[language=Rose]
def f0(y: Real): Real =
  5 / y
def f1(x: Real): Real =
  7 / y
def f2(x: Real): Real =
  5 / y
def h(z: Real): Real =
  f1(z) + f1(z)
\end{lstlisting}
\end{codeblock}
\end{subfigure}
\caption{JavaScript code (left) that produces \Rose{} IR (right) when evaluated.}
\label{fig:metaprogramming-example}
\end{figure}

We have described the right-hand side of \cref{fig:architecture}; now all that remains in this section is to describe the ``evaluate'' step on the left-hand side. \Cref{fig:metaprogramming-example} shows a simple example of how evaluating JavaScript code produces \Rose{} IR. In this example, we have two JavaScript functions \code{f} and \code{g}, each of which uses \code{fn} to construct and return a \Rose{} function when called. We call \code{f} three times and never call \code{g}, so three \Rose{} IR functions resulted from the single instance of \code{fn} in the source of \code{f}, and zero \Rose{} IR functions resulted from the instance of \code{fn} in the source of \code{g}. Then, we call \code{fn} one more time, and in the body of that \code{fn}, we call \code{f1} twice. Note that those calls to \code{f1} remain in the resulting IR; \Rose{} does not inline calls, in contrast to standard tracing autodiff frameworks.

\section{Evaluation}\label{sec:evaluation}

As discussed in \cref{sec:design}, \Rose{} is characterized by three primary design choices:
\begin{enumerate}[label=\textbf{D\arabic*}]
    \item\label{choice:fns} Allow users to define and compose custom functions using \code{fn}.
    \item\label{choice:transformation} Use program transformation to compile to WebAssembly.
    \item\label{choice:metaprogramming} Use tracing to allow metaprogramming in JavaScript.
\end{enumerate}

These design choices are motivated by \Rose's role as a toolkit for building interactive, differentiable web applications. The dynamic bandwidth- and latency-constrained environment of web browsers poses significant constraints on the size and speed of \Rose. In addition to good performance, \Rose{} also needs to be expressive and flexible enough for the end user to build web applications in a myriad of domains.

In this section, we evaluate \Rose{} both quantitatively and qualitatively by integrating \Rose{} into three web-based applications for diagramming, physical simulation, and reinforcement learning (\cref{sec:applications}). In subsequent subsections, we report on \Rose's performance (\cref{sec:performance,sec:size}) and discuss qualitative observations of how \Rose's design choices impact its expressiveness and flexibility (\cref{sec:qualitative}).

\subsection{Benchmark and applications}
\label{sec:applications}

\begin{figure}
    \centering
    \includegraphics[width=\linewidth]{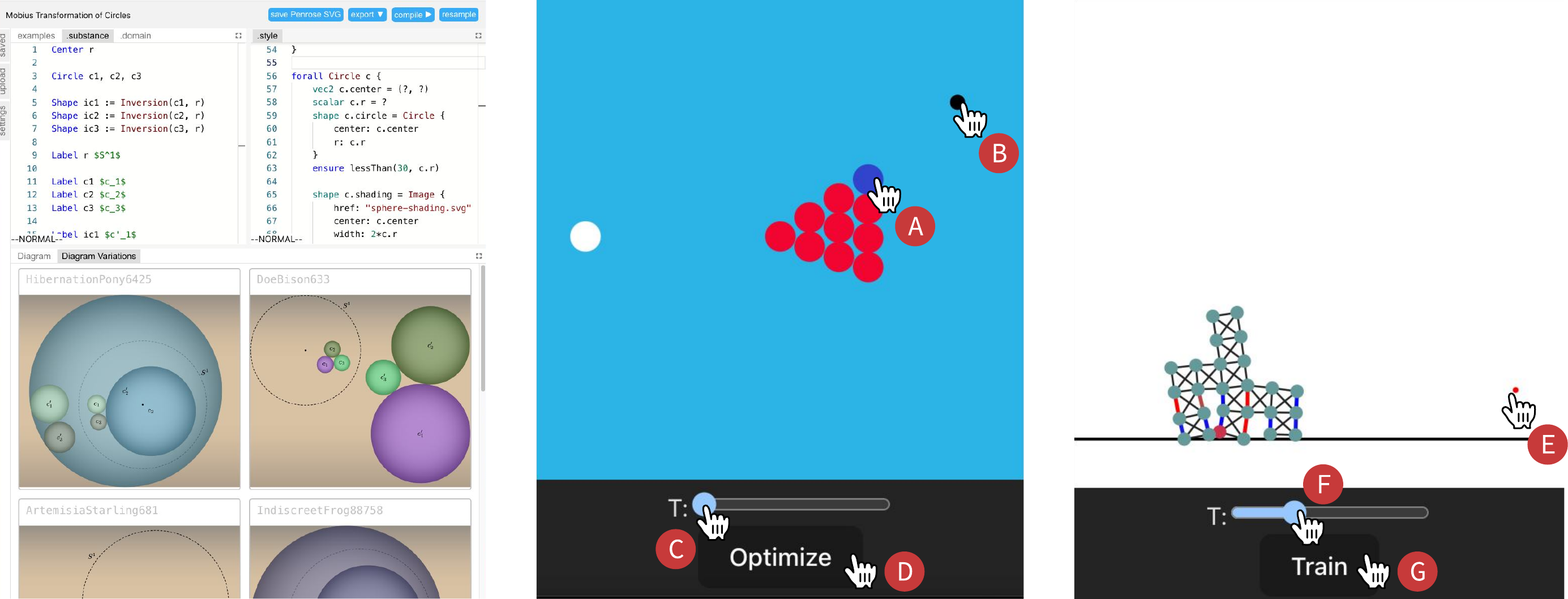}
    \caption{Three web-based applications re-implemented with \Rose. \textbf{Left:} the Penrose IDE. \textbf{Middle:} billiards simulator that optimizes \ui{D} for cue ball angle and speed such that the object ball \ui{A} reaches the target \ui{B}. \textbf{Right:} mass-spring robot controlled by a neural net trained \ui{G} with a designated goal \ui{D}. Both simulations can be replayed by dragging the sliders at any point \ui{D} and \ui{G}.}
    \label{fig:applications}
\end{figure}

To the best of our knowledge, there are no benchmark suites for evaluating autodiff performance generally~\cite{shen2021}, let alone web-based autodiff of scalar programs. Further, measuring performance on just a benchmark would limit our ability to qualitatively evaluate the expressiveness of \Rose{} in real-world applications. Therefore, we opted to find applications in which autodiff plays a central role, and re-implement the autodiff module or the entire application using \Rose. We believe this approach gives us better ecological validity (i.e.\ the realism of our evaluation setup) and potentially leads to a rich source of examples. Our search resulted in two frameworks that are rich sources of applications and benchmarks: \textbf{Penrose}~\cite{ye2020}, a web-based diagramming framework; and \textbf{Taichi}~\cite{hu2019}, a Python library for high-performance parallel programming. The two frameworks also have sufficiently different settings that together they illustrate the flexibility of \Rose.

Penrose allows the user to specify a diagram by constructing a custom numerical optimization problem in a DSL called Style, then runs a numerical solver to rearrange the shapes in the diagram until it finds a local minimum. Optimizing the layout of these diagrams involves defining and differentiating a wide range of mathematical operations on scalars, from simple operations like finding the distance between points to more sophisticated calculations like Minkowski addition, KL divergence, and silhouette points. The main application of Penrose is a web-based IDE (\cref{fig:applications}, left), where users live-edit programs to produce layout-optimized diagrams. Importantly, the framework uses TensorFlow.js for autodiff and ships with 173 ``registry'' diagrams for performance testing, each of which was compiled to a unique differentiable computation. Therefore, we deemed it as a suitable target for performance comparison with TensorFlow.js, the closest baseline to which we can compare \Rose's performance. This set of registry diagrams is quite diverse, comprising a total of 97 unique Style programs which are preprocessed by the Penrose compiler frontend before being passed to \Rose.

Many applications of Taichi involve differentiable programming and DiffTaichi~\cite{hu2020} presents a few example Python applications that combines physical simulation and neural networks. We used \Rose{} to implement and augment two such differentiable simulations from Taichi: \code{billiards} and \code{robot} (\cref{fig:applications}, middle and right). The \code{billiards} example is a differentiable simulation of pool combination shots. The program simulates rigid body collisions between a cue ball and object balls. Leveraging the differentiability of the simulation, a gradient descent optimizer solves for the initial position and velocity of the cue ball to send a designated object ball to a target position. The \code{robot} example simulates a robot made of a mass-spring system, where springs are actuated to move the robot towards a goal position. A neural network controller is trained on simulator gradients to update the spring actuation magnitude over time. In both cases, the simulation is run to completion, remembering intermediate computations along the way, and then autodiff is used to run back through the simulation in reverse to compute gradients for the initial state.

All three applications were implemented using \Rose's JavaScript binding. They all run in major browsers such as Safari and Chrome. To showcase the benefits of running in the web browser, we added interactive features to the Taichi applications (\cref{fig:applications}). For instance, the Taichi version of \code{billiards} is a command-line application that outputs a series of static images based on hard-coded parameters for the choice of the object ball and goal position. The \Rose{} version allows the user to interactively explore the simulator by selecting the object ball (\cref{fig:applications}\ui{A}), moving the goal position (\cref{fig:applications}\ui{B}), optimizing the cue ball position (\cref{fig:applications}\ui{D}), and re-playing the simulation (\cref{fig:applications}\ui{C}).

\subsection{Size}
\label{sec:size}

Similar to TensorFlow.js, \Rose{} is a client-side JavaScript package that is typically bundled with the rest of a web application and delivered over the internet. To run in web browsers, \Rose{} needs to be comparable or smaller then similar packages such as TensorFlow.js. As a baseline, a common TensorFlow.js distribution, \code{@tensorflow/tfjs-core} version 4.18.0 (latest at time of writing), is 479.98 kB after minification (85.88 kB after gzip). We publish \Rose{} via the npm package \code{rose},\footnote{\url{https://www.npmjs.com/package/rose}} which is at version 0.4.10 at time of writing. The \Rose{} WebAssembly binary size~\cite{ayers2022} is 168.51 kB (63.97 kB after gzip), and the JavaScript wrapper layer is 31.77 kB after minification (8.74 kB after gzip). For a more extreme comparison, there are projects that package heavy compiler infrastructure like LLVM to WebAssembly~\cite{soedirgo2023}, but those produce binaries on the order of a hundred megabytes, causing unacceptable load times for end users. Another reference point is the Taichi package on PyPI, which is about 50--80 megabytes depending on the platform. It is unclear how difficult it would be to package Taichi to run in the browser using Pyodide~\cite{pyodide2019}, which is 6.4 megabytes and whose authors claim to be $3\times$--$5\times$ slower than native Python.

\subsection{Performance}\label{sec:performance}

\begin{figure}
    \centering
    \includegraphics{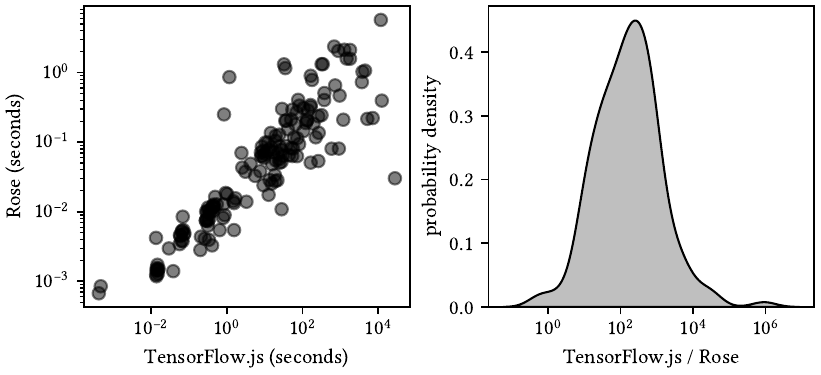}
    \caption{\textbf{Left:} Log-log scatterplot of Penrose diagram optimization time with TensorFlow.js versus \Rose. \textbf{Right:} Log-scale kernel density estimate (KDE) plot of the optimization time of TensorFlow.js to \Rose.}
    \label{fig:data}
\end{figure}

To compare with the TensorFlow.js baseline for execution performance, we replaced the Penrose TensorFlow.js-based autodiff engine with one written in \Rose{} and ran both versions on the benchmark of 173 diagrams (\cref{sec:applications}). We measured the amount of time it took for each autodiff engine to perform any necessary compilation, plus the time taken by the Penrose L-BFGS~\cite{liu1989} optimization engine to converge on each diagram. We specifically include the time it takes for \Rose{} to do autodiff, transposition, and WebAssembly compilation, despite the fact that TensorFlow.js does not have an analogous compilation step. On the surface this puts \Rose{} at a disadvantage, but fast compilation time is essential when constructing \Rose{} functions dynamically in a user-facing web application, as Penrose does.

\Cref{fig:data} shows the results.\footnote{All numbers we report in this section were measured in the V8 JavaScript engine (used in both Chrome and Node.js) on a 2020 MacBook Pro with M1 chip.}  The quartiles for the ratio of TensorFlow.js optimization time to \Rose{} optimization time were $37\times$, $173\times$, and $598\times$. These results show that \Rose{} provides an enormous advantage over TensorFlow.js (the state-of-the-art for autodiff on the web) for scalar programs like those found in Penrose diagrams. Because these numbers include both compile time and optimization time, the results demonstrate the end-to-end performance of \Rose.

We omitted 10 of the 173 diagrams from our data analysis:
\begin{itemize}
    \item \textbf{9 NaN failures:} Penrose aborts if it detects a ``not-a-number'' (NaN) value in the gradient as it is optimizing. This occurred in the TensorFlow.js version of Penrose for nine diagrams. \emph{The \Rose{} version of Penrose did not encounter NaNs for these programs.}
    \item \textbf{1 timeout:} For one diagram, we stopped the TensorFlow.js version of Penrose after it had run for over 24 hours. \emph{The \Rose{} version of Penrose took 42 milliseconds to compile and optimize this diagram.}
\end{itemize}

Tensorflow.js runs on both CPU and GPU. We used the \code{"cpu"} backend in our comparison because we found that, for scalar programs, it was faster than their GPU backend. To double-check this, we took the 88 diagrams (over half) that were quickest to run with TensorFlow.js, and also ran them with \code{@tensorflow/tfjs-node} and \code{@tensorflow/tfjs-node-gpu}, which they claim are faster than the \code{"cpu"} backend. We found that the Node backend is \emph{79\% slower} (median ratio) than the \code{"cpu"} backend, and the Node GPU backend is \emph{75\% slower} (median ratio) than the \code{"cpu"} backend. Also, those backends are unable to run in a browser, unlike the \code{"cpu"} backend, so they would be inappropriate for a direct comparison to \Rose.

As we will discuss in \cref{sec:scalar}, \Rose's ability to define separate functions in a graph (rather than just a single graph of scalar or tensor values) is crucial to producing small enough WebAssembly binaries to feed to the browser. To investigate whether WebAssembly brought significant performance gains in the first place to be worth facing that challenge, we compared against a modified version of \Rose{} which emits JavaScript code instead of WebAssembly. This experiment gave quartile slowdowns of 10\%, 49\%, and 100\% for optimization of Penrose diagrams, showing that WebAssembly provides a significant advantage over JavaScript as a compilation target for \Rose.

For the two Taichi applications (\cref{fig:applications}, middle and right), we compare the running time of training/optimizing with the Python command-line counterparts. On average (of 10 runs), \Rose{} is on par with the native performance of the Python versions: for the default initial condition in \code{billiards}, \Rose{} completed the optimization in $22.7\text{s} \pm 0.2\text{s}$ while Taichi completed it in $20.6\text{s} \pm 0.8\text{s}$; for the default condition in \code{robot}, \Rose{} finished 100 iterations of learning in $32.1\text{s} \pm 0.2\text{s}$ while Taichi took $31.3\text{s} \pm 1.1\text{s}$.

In the Penrose IDE (\cref{fig:applications}, left), the main interaction that will trigger differentiation is compiling the DSL to diagrams. As reported in the previous section, \Rose-based Penrose is significantly faster than the TensorFlow.js version, often leading to visible reduction in diagram layout optimization time in the user interface.

\subsection{Qualitative observations}\label{sec:qualitative}

Our implementation effort to port both Penrose and Taichi applications to \Rose{} spanned thousands of lines of code, including replacing the Penrose autodiff engine and function library and rewriting both \code{billiards} and \code{robot} for scratch. In this process, we have written a wide variety of differentiable programs using \Rose, and had a chance to observe how \Rose's main design choices (\cref{choice:fns,choice:metaprogramming,choice:transformation}) impact the way programs are written. In this section, we report on our qualitative observations using \Rose{} in these real-world settings, highlighting how these design choices interact with each other to form a coherent system.

\subsubsection{Writing scalar programs as composable functions}\label{sec:scalar}

The original versions of Penrose, \texttt{billiards}, and \texttt{robot} are naturally written as scalar programs. In Penrose, \texttt{bboxCircle} (line~\ref{line:bbox_circle} of \cref{lst:bbox}) computes the bounding box by performing arithmetic on scalar values for the center and radius of a circle. In Taichi, both \texttt{billiards} and \texttt{robot} involve hand-crafted scalar programs for differentiable simulations. For instance, \texttt{apply\_spring\_force} (\cref{fig:difftaichi-v-rose}) loops through individual springs in the robot, computing the force on the spring based on scalar-valued parameters, and scatter forces to end points of springs.

Because \Rose{} is designed for writing scalar programs, translating both Penrose and Taichi source programs to \Rose{} is straightforward and largely preserves the structures of the programs. For instance, when translating the Python programs from Taichi into TypeScript and \Rose, as shown in \cref{fig:difftaichi-v-rose}, Taichi kernels can be translated one-to-one to \Rose{} functions.

Reflecting on the design choices, the combination of transformation to WebAssembly (\cref{choice:transformation}) and the basic building block of composable functions (\cref{choice:fns}) gives the user both performance gains and an ergonomic programming interface. In the case of Taichi, the \Rose{} abstraction of \texttt{fn} is not only useful for one-to-one translation from Taichi, but also necessary for running the simulator in browsers. Major WebAssembly engines have limits on WebAssembly binary size and on the number of local variables in each function. While it is possible to encapsulate much of the simulation code of \texttt{billiards} and \texttt{robot} in bigger JavaScript functions, the compiled size and local counts of these functions would quickly exceed these limits and would not run in the browser. Therefore, segmenting the source into functional units of \texttt{fn}s effectively reduces the size of emitted WebAssembly functions and modules, avoiding these errors and reducing compile times.

\begin{figure}[!h]
\vspace{2em}
 \begin{minipage}{0.48\linewidth}
  \begin{minted}[fontsize=\scriptsize]{python}
@ti.kernel
def apply_spring_force(t: ti.i32):
    for i in range(n_springs):
        a = spring_anchor_a[i]
        b = spring_anchor_b[i]
        pos_a = x[t, a]
        pos_b = x[t, b]
        dist = pos_a - pos_b
        length = dist.norm() + 1e-4
        target_length = spring_length[i] *
            (1.0 + spring_actuation[i] * act[t, i])
        impulse = dt * (length - target_length) *
            spring_stiffness[i] / length * dist

        ti.atomic_add(v_inc[t + 1, a], -impulse)
        ti.atomic_add(v_inc[t + 1, b], impulse)
  \end{minted}
 \end{minipage}
 \begin{minipage}{0.48\linewidth}
  \begin{minted}[fontsize=\scriptsize]{js}
const apply_spring_force = fn(
  [Objects, Act], Objects, (x, act) => {
      const v_inc = [];
      for (let i = 0; i < n_objects; i++)
        v_inc.push([0, 0]);
      for (let i = 0; i < n_springs; i++) {
        const spring = robot.springs[i];
        const a = spring.object1;
        const b = spring.object2;
        const pos_a = x[a];
        const pos_b = x[b];
        const dist = vsub2(pos_a, pos_b);
        const length = add(norm(dist), 1e-4);
        const target_length = mul(spring.length,
          add(1, mul(act[i], spring.actuation))
        );
        const impulse =
          vmul(div(mul(dt * spring.stiffness,
                       sub(length, target_length)),
                   length),
               dist);
        v_inc[a] = vsub2(v_inc[a], impulse);
        v_inc[b] = vadd2(v_inc[b], impulse);
      }
      return v_inc;
});
  \end{minted}
 \end{minipage}
\caption{A function that applies spring actuation on the mass-spring robot model in the \texttt{robot} example, written in Taichi (\textbf{Left}) and \Rose{} (\textbf{Right}). The translation from Taichi to \Rose{} is straightforward.}
\label{fig:difftaichi-v-rose}
\end{figure}

\subsubsection{Metaprogramming and function dynamism}\label{sec:dynamism}

The \Rose{} IR is designed to be performant and easy to compile to WebAssembly (\cref{choice:transformation}) and therefore has limited expressiveness (\cref{sec:design}). Metaprogramming using JavaScript enables the user to dynamically generate complex computation graphs that are impossible to specify with the \Rose{} IR alone (\cref{choice:metaprogramming}). For instance, the \texttt{bboxGroup} function in \cref{lst:bbox} computes the bounding box of a \texttt{Group} in Penrose, a recursive collection of shapes. For non-collection shape types such as \texttt{Circle}, we ported the TensorFlow.js implementation to \Rose{} easily, e.g.\ \texttt{bboxCircle}. However, \texttt{bboxGroup} needs to recurse over the \texttt{Group} data structure to find out the bounding boxes of individual shapes before aggregating them into the final bounding box. This requires conditional dispatch of (1) \Rose{} functions based on a discrete tag (\texttt{shape.kind}) and (2) recursive calls to \texttt{bboxGroup} to handle nested groups.

\begin{listing}[t]
\begin{codeblock}
\begin{minted}[linenos,escapeinside=||,]{js}
const bboxGroup = (shapes) => {
  const bboxes = shapes.map(bbox);
  const left = bboxes.map((b) => b.left).reduce(min);
  const right = bboxes.map((b) => b.right).reduce(max);
  const bottom = bboxes.map((b) => b.bottom).reduce(min);
  const top = bboxes.map((b) => b.top).reduce(max);
  return { left, right, bottom, top };
};

const bboxCircle = fn([Circle], Rectangle,|\label{line:bbox_circle}|
  ({ center: [x, y], radius: r }) => {
    const left = sub(x, r);
    const right = add(x, r);
    const bottom = sub(y, r);
    const top = add(y, r);
    return { left, right, bottom, top };
  },
);

const bbox = (shape) => {
  switch (shape.kind) {
    case "Rectangle": return shape.value;
    case "Circle": return bboxCircle(shape.value);
    case "Group": return bboxGroup(shape.value);
  }
};
\end{minted}
\end{codeblock}
\caption{Examples of JavaScript metaprogramming to construct \Rose{} functions for recursive data structures.}
\label{lst:bbox}
\end{listing}

\begin{figure}[t]
    \centering
    \includegraphics[width=.7\linewidth]{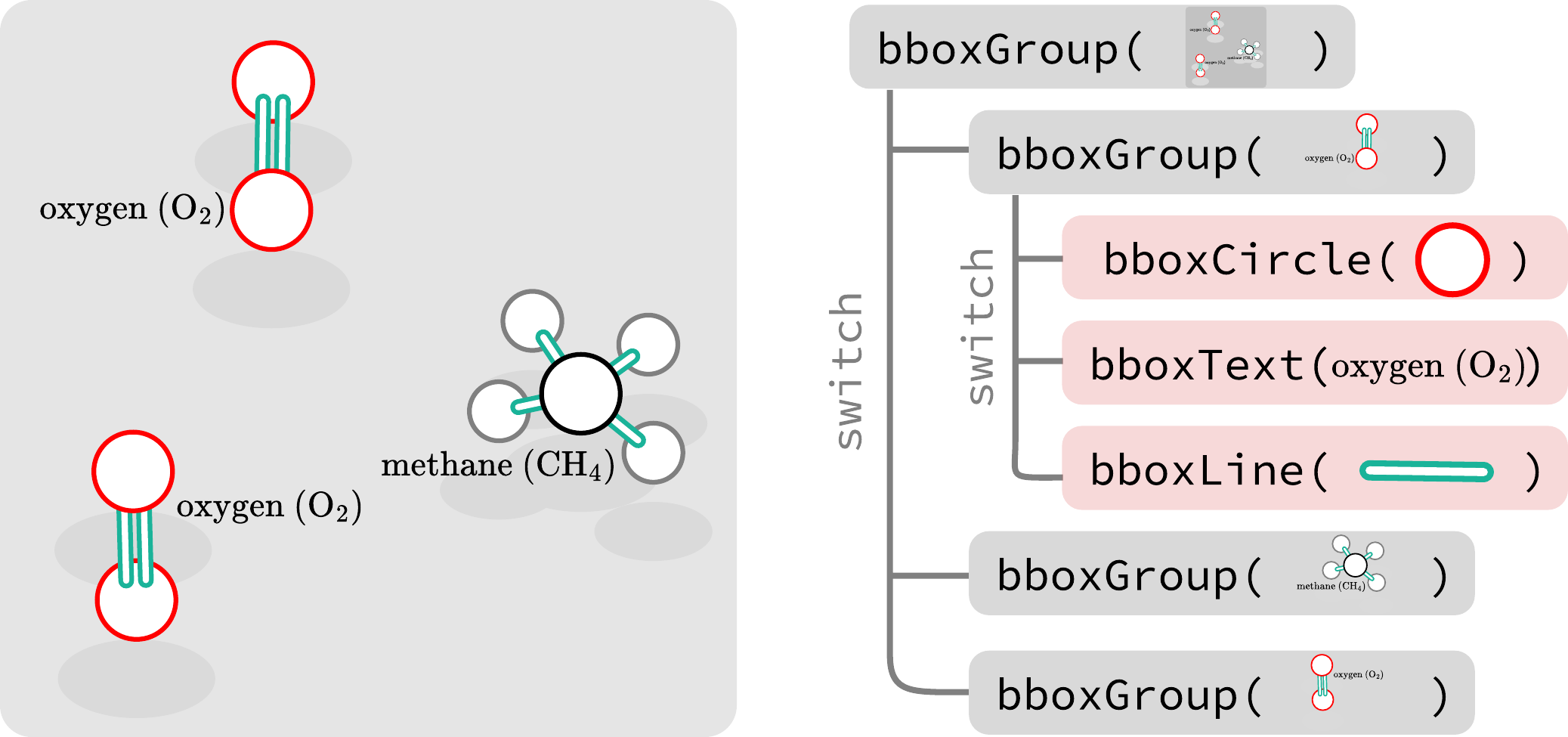}
    \caption{In Penrose, we used JavaScript to programmatically generate \Rose{} functions. \textbf{Left:} a figure comprised of a top-level group containing all molecules and sub-groups for each molecule. \textbf{Right:} the \texttt{bboxGroup} function conditionally generates \Rose{} functions or recursively calls itself based on the shape type.}
    \label{fig:bbox-group}
\end{figure}

\Cref{fig:bbox-group} shows an example of calling \texttt{bboxGroup} on nested groups of shapes. The diagram in \cref{fig:bbox-group} (left) has 1 group containing the whole diagram, and 3 subgroups of molecules that contain shapes such as \texttt{Text} and \texttt{Circle}. \Cref{fig:bbox-group} shows how \texttt{bboxGroup} uses JavaScript language features to compose \Rose{} functions into a computation graph, denoting JavaScript constructs in gray and \Rose{} functions in red. First, for each member shape, we \texttt{switch} on \texttt{shape.kind} to determine whether to (a) call individual \Rose{} bounding box functions like \texttt{bboxCircle} or (b) recurse to call \texttt{bboxGroup}. Then, after all the child bounding boxes are computed, we use JavaScript \texttt{map} and \texttt{reduce} to aggregate via \Rose{} \texttt{min} and \texttt{max} functions.

In the case of Penrose, metaprogramming actually helped us reduce the lines of code to refactor, because many plain JavaScript functions can stay the same and we only had to refactor functions that involve actual computation. We also speculate that by reducing the size of \Rose-specific constructs, new users can learn a smaller API easier and experience a smoother learning curve.

\section{Related work}\label{sec:related}

Autodiff first started being seriously studied a few decades ago~\cite{speelpenning1980}, with Griewank and Walther's book~\cite{griewank2008} consolidating research on the topic up until its publication. Some tools were developed, such as Tapenade~\cite{hascoet2013} which operates over C and Fortran. The machine learning community developed an interest in autodiff over the past decade, resulting in popular tools including TensorFlow~\cite{abadi2016}, PyTorch~\cite{paszke2019}, and JAX~\cite{frostig2018} as mentioned in \cref{sec:introduction}. JAX is particularly interesting because in a way it blends together tracing and transformation like we do here, but unlike \Rose, JAX is not scalar-friendly and does not allow the programmer to explicitly define functions to serve as boundaries for tracing. Other autodiff systems include Zygote~\cite{innes2019zygote} for Julia, and Enzyme~\cite{moses2020,moses2021,moses2022} for LLVM IR~\cite{lattner2004}. For graphics programming, A$\delta$~\cite{yang2022} and Dr.Jit~\cite{jakob2022} can be used to differentiate shaders.

The programming languages community has also taken an interest in autodiff~\cite{pearlmutter2008}, producing proofs of correctness~\cite{abadi2019}, program transformations for SSA~\cite{innes2019ssa} and CPS~\cite{wang2019}, and more recently, reformulations of reverse-mode autodiff in terms of dual numbers~\cite{smeding2023}, as well as a new autodiff formulation called CHAD~\cite{vakar2022,smeding2024}. Some work also attempts to bridge the gap between programming language theory and the machine learning domain by facilitating automatic parallelization~\cite{bernstein2020,paszke2021}. The latter work also resulted in a formalization of function transposition~\cite{radul2023} which directly inspired the low-level design of autodiff in \Rose.

\Rose{} supports higher-order derivatives because its core IR is closed under differentiation and transposition. A more sophisticated approach we don't explore here would be derivative towers~\cite{karczmarczuk1998,pearlmutter2007}, sometimes called ``Taylor towers'' because they use Taylor expansions instead of the chain rule. We would be interested to see how derivative towers can be combined with our approach in future work, while avoiding the pitfalls of perturbation confusion~\cite{siskind2008,manzyuk2019}. Another optimization that becomes crucial when scaling up autodiff is checkpointing, which cuts down drastically on memory requirements; for instance, while the semantics of \Rose{} can in general result in keeping around arbitrarily many intermediate results, a recursive ``divide-and-conquer'' checkpointing scheme~\cite{siskind2018} reduces the memory impact of reverse-mode autodiff to a logarithmic factor; the cost, though, is that the asymptotic running time would also suffer a logarithmic factor.

\section{Conclusion and future work}\label{sec:conclusion}

This paper introduced \Rose, an embedded domain-specific language for automatic differentiation of interactive programs on the web, which blends together the two primary autodiff techniques of tracing and transformation. Currently \Rose{} targets WebAssembly, which runs on the CPU; as we showed in \cref{sec:performance}, this already provides an enormous performance advantage for scalar programs when compared to the state-of-the-art for autodiff on the web. In future work, we would also like to pursue further performance gains by implementing a backend that targets WebGPU~\cite{kenwright2022}. We have already laid the groundwork for this by drawing inspiration for the \Rose{} IR from Dex~\cite{paszke2021} to be friendly to automatic parallelization, such as the \code{for} construct and accumulate-only reference types. In general, we plan to continue this line of work to open up new modes of differentiable interactivity.

\bibliography{refs}

\appendix

\section{Inference rules}\label{sec:inference}

This appendix includes \cref{fig:typing-expr,fig:typing-blocks} from \cref{sec:ir}, \cref{fig:forward-mode-type-expr,fig:forward-mode-block} from \cref{sec:autodiff}, and \cref{fig:transpose-expr,fig:transpose-block} from \cref{sec:transpose}. \Cref{fig:transpose-block} makes use of one helper function:
\begin{align*}
    \tuple(x, \epsilon) &= \pair{x}{\unit} \\
    \tuple(x, \cons{\epsilon}{y}) &= \pair{x}{\pair{y}{\unit}} \\
    \tuple(x, \cons{\cons{T}{y}}{z}) &= \bindinfer{w}{\pair{y}{z}}{\tuple(x, T :: w)}
\end{align*}

\begin{figure}
\begin{gather*}
    \fbox{$\underline{\tau : \kappa}, \underline{x : \tau} \vdash e : \tau$}
    \quad
    \inference{}{\vdash \unit : \Unit}
    \quad
    \inference{}{\vdash \true : \Bool}
    \quad
    \inference{}{\vdash \false : \Bool}
    \infline
    \inference{}{\vdash c : \Real}
    \quad
    \inference{m < n}{\vdash m : n}
    \quad
    \inference{|\underline{x}| = n}{\underline{x : \tau} \vdash \arr{\underline{x}} : \Arr{n}{\tau}}
    \infline
    \inference{}{x : \tau, y : \tau' \vdash \pair{x}{y} : \Prod{\tau}{\tau'}}
    \quad
    \inference{}{x : \Bool \vdash \lnot x : \Bool}
    \infline
    \inference{\ominus \in \{-, \abs, \sgn, \ceil, \floor, \trunc, \code{sqrt}\}}{x : \Real \vdash \ominus x : \Real}
    \quad
    \inference{\oplus \in \{+, -, \times, \div\}}{x : \Real, y : \Real \vdash x \oplus y : \Real}
    \infline
    \inference{\oplus \in \{\land, \lor, \xnor, \xor\}}{x : \Bool, y : \Bool \vdash x \oplus y : \Bool}
    \quad
    \inference{\oplus \in \{<, \leq, =, >, \geq\}}{x : \Real, y : \Real \vdash x \oplus y : \Bool}
    \infline
    \inference{}{x : \Bool, y : \tau, z: \tau \vdash \tern{x}{y}{z} : \tau}
    \quad
    \inference{}{x : \Acc{\tau}, y : \tau \vdash \acc{x}{y} : \Unit}
    \infline
    \inference{}{x : \Arr{\tau}{\tau'}, y : \tau \vdash \elem{x}{y} : \tau'}
    \quad
    \inference{}{x : \Acc{\Arr{\tau}{\tau'}}, y : \tau \vdash \slice{x}{y} : \Acc{\tau'}}
    \infline
    \inference{}{x : \Prod{\tau}{\tau'} \vdash \fst{x} : \tau}
    \quad
    \inference{}{x : \Prod{\tau}{\tau'} \vdash \snd{x} : \tau'}
    \infline
    \quad
    \inference{}{x : \Acc{\Prod{\tau}{\tau'}} \vdash \rfst{x} : \Acc{\tau}}
    \quad
    \inference{}{x : \Acc{\Prod{\tau}{\tau'}} \vdash \rsnd{x} : \Acc{\tau'}}
    \infline
    \inference{\Gamma \vdash \underline{\tau' : \kappa}}{\Gamma, f : \fty{\underline{\decl{t}{\kappa}}}{\underline{\tau}}{\tau'}, \underline{x : \tau[\overline{t / \tau''}]} \vdash \call{f}{\underline{\tau''}}{\underline{x}} : \tau'[\underline{t / \tau''}]}
    \infline
    \inference{\Gamma, x : \tau \vdash b : \tau'}{\Gamma \vdash \map{x}{\tau}{b} : \Arr{\tau}{\tau'}}
    \quad
    \inference{\Gamma, y: \tau, x: \Acc{\tau} \vdash b : \tau'}{\Gamma, y : \tau \vdash \accum{x}{y}{b} : \Prod{\tau}{\tau'}}
\end{gather*}
\caption{Typing rules for \Rose{} expressions.}
\label{fig:typing-expr}
\end{figure}

\begin{figure}
\begin{gather*}
    \fbox{$\underline{\tau : \kappa} \vdash \tau : \kappa$}
    \quad
    \inference{}{t : \kappa \vdash t : \kappa}
    \quad
    \inference{\Gamma \vdash \tau : \Value}{\Gamma \vdash \tau : \Type}
    \quad
    \inference{\Gamma \vdash \tau : \Index}{\Gamma \vdash \tau : \Value}
    \infline
    \inference{}{\vdash \Unit : \Value}
    \quad
    \inference{}{\vdash \Bool : \Value}
    \quad
    \inference{}{\vdash \Real : \Value}
    \infline
    \inference{}{\vdash n : \Index}
    \quad
    \inference{\Gamma \vdash \tau : \Value}{\Gamma \vdash \Acc{\tau} : \Type}
    \infline
    \inference{
        \Gamma \vdash \tau : \Index
        &
        \Gamma \vdash \tau' : \Value
    }{\Gamma \vdash \Arr{\tau}{\tau'} : \Value}
    \quad
    \inference{
        \Gamma \vdash \tau : \Value
        &
        \Gamma \vdash \tau' : \Value
    }{\Gamma \vdash \Prod{\tau}{\tau'} : \Value}
    \infline
    \fbox{$\underline{x : \tau} \vdash b : \tau$}
    \quad
    \inference{}{x : \tau \vdash x : \tau}
    \quad
    \inference{
        \Gamma \vdash \tau : \Type
        &
        \Gamma \vdash e : \tau
        &
        \Gamma, x : \tau \vdash b : \tau'
    }{\Gamma \vdash \bind{x}{\tau}{e}{b} : \tau'}
    \infline
    \fbox{$\underline{x : \tau} \vdash \underline{d}$}
    \quad
    \inference{}{\vdash}
    \infline
    \inference{
        \Gamma, \underline{t : \kappa} \vdash \tau' : \Value
        &
        \Gamma, \underline{t : \kappa}, \underline{x : \tau} \vdash b : \tau'
        &
        \Gamma, f : \fty{\underline{\decl{t}{\kappa}}}{\underline{\tau}}{\tau'} \vdash \underline{d}
    }{\Gamma \vdash \fn{f}{\underline{\decl{t}{\kappa}}}{\underline{\decl{x}{\tau}}}{\tau'}{b}\codespace\underline{d}}
\end{gather*}
\caption{Typing rules for \Rose{} blocks and function definitions.}
\label{fig:typing-blocks}
\end{figure}

\begin{figure}
\begin{gather*}
    \fbox{$\autodifftype{\tau}{\tau}$}
    \quad
    \inference{}{\autodifftype{\Unit}{\Unit}}
    \quad
    \inference{}{\autodifftype{\Bool}{\Bool}}
    \quad
    \inference{}{\autodifftype{n}{n}}
    \quad
    \inference{}{\autodifftype{t}{t}}
    \infline
    \inference{\autodifftype{\tau}{\tau'}}{\autodifftype{\Acc{\tau}}{\Acc{\tau'}}}
    \quad
    \inference{
        \autodifftype{\tau_1}{\tau_1'}
        &
        \autodifftype{\tau_2}{\tau_2'}
    }{
        \autodifftype{\Arr{\tau_1}{\tau_2}}{\Arr{\tau_1'}{\tau_2'}}
    }
    \quad
    \inference{
        \autodifftype{\tau_1}{\tau_1'}
        &
        \autodifftype{\tau_2}{\tau_2'}
    }{
        \autodifftype{\Prod{\tau_1}{\tau_2}}{\Prod{\tau_1'}{\tau_2'}}
    }
    \infline
    \inference{}{\autodifftype{\Real}{\Prod{\Real}{\Real}}}
    \infline
    \fbox{$\autodiffexpr{e}{e}$}
    \quad
    \inference{}{\autodiffexpr{\unit}{\unit}}
    \quad
    \inference{}{\autodiffexpr{\true}{\true}}
    \quad
    \inference{}{\autodiffexpr{\false}{\false}}
    \infline
    \inference{}{\autodiffexpr{n}{n}}
    \quad
    \inference{}{\autodiffexpr{\arr{\underline{x}}}{\arr{\underline{x}}}}
    \quad
    \inference{}{\autodiffexpr{\pair{x}{y}}{\pair{x}{y}}}
    \infline
    \inference{}{\autodiffexpr{\elem{a}{i}}{\elem{a}{i}}}
    \quad
    \inference{}{\autodiffexpr{\fst{\pi}}{\fst{\pi}}}
    \quad
    \inference{}{\autodiffexpr{\snd{\pi}}{\snd{\pi}}}
    \infline
    \inference{}{\autodiffexpr{\slice{a}{i}}{\slice{a}{i}}}
    \quad
    \inference{}{\autodiffexpr{\rfst{\pi}}{\rfst{\pi}}}
    \quad
    \inference{}{\autodiffexpr{\rsnd{\pi}}{\rsnd{\pi}}}
    \infline
    \inference{}{\autodiffexpr{\tern{p}{x}{y}}{\tern{p}{x}{y}}}
    \quad
    \inference{}{\autodiffexpr{\acc{x}{y}}{\acc{x}{y}}}
    \quad
    \inference{}{\autodiffexpr{\lnot x}{\lnot x}}
    \infline
    \inference{
        \oplus \in \{\land, \lor, \xnor, \xor, \neq, <, \leq, =, >, \geq\}
    }{
        \autodiffexpr{x \oplus y}{x \oplus y}
    }
    \quad
    \inference{
        \autodiffblock{b}{b'}
        &
        b'' := \accum{x}{y}{b'}
    }{
        \autodiffexpr{\accum{x}{y}{b}}{b''}
    }
    \infline
    \inference{
        \autodifftype{\tau}{\tau'}
        &
        \autodiffblock{b}{b'}
    }{
        \autodiffexpr{\map{i}{\tau}{b}}{\map{i}{\tau'}{b'}}
    }
    \quad
    \inference{
        \underline{\autodifftype{\tau}{\tau'}}
    }{
        \autodiffexpr{
            \call{f}{\underline{\tau}}{\underline{x}}
        }{
            \call{f'}{\underline{\tau'}}{\underline{x}}
        }
    }
\end{gather*}
\caption{Inference rules for forward-mode autodiff of types and some expressions.}
\label{fig:forward-mode-type-expr}
\end{figure}

\begin{figure}
\begin{gather*}
    \fbox{$\autodiffunary{x}{\ominus}{e}{e}$}
    \quad
    \inference{
        \ominus \in \{\sgn, \ceil, \floor, \trunc\}
    }{
        \autodiffunary{x}{\ominus}{\ominus x}{0}
    }
    \infline
    \inference{}{\autodiffunary{x}{\code{-}}{\code{-}x}{\code{-}\hat{x}}}
    \quad
    \inference{}{\autodiffunary{x}{\abs}{\abs\codespace x}{\hat{x} \times \sgn\codespace x}}
    \infline
    \fbox{$\autodiffbinary{x}{y}{\oplus}{e}{e}$}
    \quad
    \inference{}{\autodiffbinary{x}{y}{+}{x + y}{\hat{x} + \hat{y}}}
    \quad
    \inference{}{\autodiffbinary{x}{y}{-}{x - y}{\hat{x} - \hat{y}}}
    \infline
    \inference{}{
        \autodiffbinary{x}{y}{\times}{x \times y}{
            \paren{\hat{x} \times y} + \paren{\hat{y} \times x}
        }
    }
    \infline
    \inference{}{
        \autodiffbinary{x}{y}{\div}{x \div y}{
            \paren{\paren{\hat{x} \times y} - \paren{\hat{y} \times x}} \div \paren{y \times y}
        }
    }
    \infline
    \fbox{$\autodiffblock{b}{b}$}
    \quad
    \inference{}{\autodiffblock{x}{x}}
    \quad
    \inference{
        \autodifftype{\tau}{\tau'}
        &
        \autodiffexpr{e}{e'}
        &
        \autodiffblock{b}{b'}
    }{
        \autodiffblock{\bind{x}{\tau}{e}{b}}{\bind{x}{\tau'}{e'}{b'}}
    }
    \infline
    \inference{
        \autodiffblock{b}{b'}
    }{
        \autodiffblock{\bindinfer{x}{c}{b}}{\bindinfer{x}{\pair{c}{0}}{b'}}
    }
    \infline
    \inference{
        \autodiffunary{x}{\ominus}{e}{e'}
        &
        \autodiffblock{b}{b'}
    }{
        \autodiffblock{\bindinfer{y}{\ominus \pi}{b}}{
            \bindinfer{\pair{x}{\hat{x}}}{\pi}{
                \bindinfer{y}{\pair{e}{e'}}{b'}
            }
        }
    }
    \infline
    \inference{
        \autodiffbinary{x}{y}{\oplus}{e}{e'}
        &
        \autodiffblock{b}{b'}
        &
        b'' := \bindinfer{y}{\pair{e}{e'}}{b'}
    }{
        \autodiffblock{\bindinfer{z}{\pi \oplus \pi'}{b}}{
            \bindinfer{\pair{\pair{x}{\hat{x}}}{\pair{y}{\hat{y}}}}{\pair{\pi}{\pi'}}{b''}
        }
    }
\end{gather*}
\caption{Inference rules for forward-mode autodiff of operators and blocks.}
\label{fig:forward-mode-block}
\end{figure}

\begin{figure}
\begin{gather*}
    \fbox{$\transposexpr{x}{e}{e}{b}$}
    \quad
    \inference{
        b := \sequence{
            \acc{\ddot{x}_0}{\elem{\dot{a}}{0}}
        }{
            \sequence{\dots}{
                \acc{\ddot{x}_{n-1}}{\elem{\dot{a}}{n-1}}
            }
        }
    }{
        \transposexpr{a}{\arr{x_0\comma \dots\comma x_{n-1}}}{
            \arr{x_0\comma \dots\comma x_{n-1}}
        }{b}
    }
    \infline
    \inference{}{\transposexpr{u}{\unit}{\unit}{\unit}}
    \quad
    \inference{}{\transposexpr{p}{\true}{\true}{\unit}}
    \quad
    \inference{}{\transposexpr{p}{\false}{\false}{\unit}}
    \infline
    \inference{}{\transposexpr{x}{c}{c}{\unit}}
    \quad
    \inference{}{\transposexpr{a}{n}{n}{\unit}}
    \quad
    \inference{}{
        \transposexpr{u}{\acc{\ddot{x}}{y}}{\acc{\ddot{x}}{y}}{\acc{\ddot{y}}{\dot{x}}}
    }
    \infline
    \inference{
        b := \bindinfer{\pair{\dot{x}}{\dot{y}}}{\dot{\pi}}{
            \sequence{\acc{\ddot{x}}{\dot{x}}}{\acc{\ddot{y}}{\dot{y}}}
        }
    }{
        \transposexpr{\pi}{\pair{x}{y}}{\pair{x}{y}}{b}
    }
    \quad
    \inference{
        b := \bindinfer{\ddot{w}}{\tern{p}{\ddot{x}}{\ddot{y}}}{\acc{\ddot{w}}{\dot{z}}}
    }{
        \transposexpr{z}{\tern{p}{x}{y}}{\tern{p}{x}{y}}{b}
    }
    \infline
    \inference{}{\transposexpr{y}{\ominus x}{\ominus x}{\unit}}
    \quad
    \inference{}{\transposexpr{z}{x \oplus y}{x \oplus y}{\unit}}
    \quad
    \inference{}{
        \transposexpr{\hat{y}}{\code{-}\hat{x}}{0}{\acc{\ddot{\hat{x}}}{\code{-}\dot{\hat{y}}}}
    }
    \infline
    \inference{}{
        \transposexpr{\hat{z}}{\hat{x} + \hat{y}}{0}{
            \sequence{\acc{\ddot{\hat{x}}}{\dot{\hat{z}}}}{\acc{\ddot{\hat{y}}}{\dot{\hat{z}}}}
        }
    }
    \quad
    \inference{}{
        \transposexpr{\hat{z}}{\hat{x} \times y}{0}{
            \acc{\ddot{\hat{x}}}{\dot{\hat{z}} \times y}
        }
    }
    \infline
    \inference{}{
        \transposexpr{\hat{z}}{\hat{x} - \hat{y}}{0}{
            \sequence{\acc{\ddot{\hat{x}}}{\dot{\hat{z}}}}{\acc{\ddot{\hat{y}}}{\code{-}\dot{\hat{z}}}}
        }
    }
    \quad
    \inference{}{
        \transposexpr{\hat{z}}{\hat{x} \div y}{0}{
            \acc{\ddot{\hat{x}}}{\dot{\hat{z}} \div y}
        }
    }
\end{gather*}
\caption{Inference rules for transposition of some expressions.}
\label{fig:transpose-expr}
\end{figure}

\begin{figure}
\begin{gather*}
    \fbox{$\underline{\tau : \kappa} \vdash \transposeblock{x}{x}{\underline{x}}{b}{b}{b}$}
    \quad
    \inference{}{
        \Gamma \vdash \transposeblock{y}{t}{T}{x}{
            \tuple(x, T)
        }{
            \acc{\ddot{x}}{\dot{y}}
        }
    }
    \infline
    \inference{
        \Gamma \vdash \tau : \Value
        &
        \transposexpr{x}{e}{e'}{b_1}
        &
        \Gamma \vdash \transposeblock{y}{t'}{\cons{T}{x}}{b}{b_2}{b_3}
        \\
        b_4 := \bindinfer{\pair{x}{t'}}{t}{
            \bindinfer{\pair{\dot{x}}{\unit}}{
                \paren{\accum{\ddot{x}}{x}{b_3}}
            }{b_1}
        }
    }{
        \Gamma \vdash \transposeblock{y}{t}{T}{
            \bind{x}{\tau}{e}{b}
        }{
            \bindinfer{x}{e'}{b_2}
        }{b_4}
    }
    \infline
    \inference{
        \Gamma \vdash \transposeblock{w}{t}{T}{b}{b_1}{b_2}
        &
        b_3 := \bindinfer{\dot{z}}{\tern{p}{\dot{x}}{\dot{y}}}{b_2}
    }{
        \Gamma \vdash \transposeblock{w}{t}{T}{
            \bindinfer{\ddot{z}}{\tern{p}{\ddot{x}}{\ddot{y}}}{b}
        }{
            \bindinfer{\ddot{z}}{\tern{p}{\ddot{x}}{\ddot{y}}}{b_1}
        }{b_3}
    }
    \infline
    \inference{
        \Gamma \vdash \transposeblock{y}{t}{T}{b}{b_1}{b_2}
        &
        b_3 := \bindinfer{x}{\elem{a}{i}}{
            \bindinfer{\ddot{x}}{\slice{\ddot{a}}{i}}{b_2}
        }
    }{
        \Gamma \vdash \transposeblock{y}{t}{T}{
            \bindinfer{x}{\elem{a}{i}}{b}
        }{
            \bindinfer{x}{\elem{a}{i}}{b_1}
        }{b_3}
    }
    \infline
    \inference{
        f \in \{\code{fst}, \code{snd}\}
        &
        \Gamma \vdash \transposeblock{y}{t}{T}{b}{b_1}{b_2}
        \\
        b_3 := \bindinfer{x}{f(\pi)}{
            \bindinfer{\ddot{x}}{\code{\&}f(\ddot{\pi})}{b_2}
        }
    }{
        \Gamma \vdash \transposeblock{y}{t}{T}{
            \bindinfer{x}{f(\pi)}{b}
        }{
            \bindinfer{x}{f(\pi)}{b_1}
        }{b_3}
    }
    \infline
    \inference{
        f \in \{\code{fst}, \code{snd}\}
        &
        \Gamma \vdash \transposeblock{y}{t}{T}{b}{b_1}{b_2}
        &
        b_3 := \bindinfer{\dot{x}}{f(\dot{\pi})}{b_2}
    }{
        \Gamma \vdash \transposeblock{y}{t}{T}{
            \bindinfer{\ddot{x}}{\code{\&}f(\pi)}{b}
        }{
            \bindinfer{\ddot{x}}{\code{\&}f(\pi)}{b_1}
        }{b_3}
    }
    \infline
    \inference{
        \Gamma \vdash \transposeblock{y}{t''}{\cons{\cons{T}{y}}{t'}}{b}{b_1}{b_2}
        &
        e := \accum{\ddot{y}}{y}{b_2}
        \\
        b_3 := \bindinfer{\pair{y}{\pair{t'}{t''}}}{t}{
            \bindinfer{\pair{\dot{y}}{\unit}}{e}{
                \call{f'''}{\underline{\tau}}{\underline{\ddot{x}}\comma \dot{y}\comma t'}
            }
        }
    }{
        \Gamma \vdash \transposeblock{z}{t}{T}{
            \bindinfer{y}{\call{f}{\underline{\tau}}{\underline{x}}}{b}
        }{
            \bindinfer{\pair{y}{t}}{\call{f''}{\underline{\tau}}{\underline{x}}}{b_1}
        }{b_3}
    }
    \infline
    \inference{
        \Gamma \vdash \transposeblock{y}{t'}{\epsilon}{b}{b_2}{b_3}
        &
        \Gamma \vdash \transposeblock{z}{t'''}{\cons{\cons{T}{a}}{t''}}{b_1}{b_4}{b_5}
        \\
        e := \map{i}{\tau}{b_2}
        &
        e' := \map{i}{\tau}{\fst{\elem{v}{i}}}
        &
        e'' := \map{i}{\tau}{\snd{\elem{v}{i}}}
        \\
        b_6 := \bindinfer{v}{e}{
            \bindinfer{a}{e'}{
                \bindinfer{t''}{e''}{b_4}
            }
        }
        \\
        b_7 := \bindinfer{u}{
            \map{i}{\tau}{\bindinfer{t'}{\elem{t''}{i}}{b_3}}
        }{\unit}
        \\
        b_8 := \bindinfer{\pair{a}{\pair{t''}{t'''}}}{t}{
            \bindinfer{\pair{\dot{a}}{\unit}}{\paren{\accum{\ddot{a}}{a}{b_5}}}{b_7}
        }
    }{
        \Gamma \vdash \transposeblock{z}{t}{T}{\bindinfer{a}{\map{i}{\tau}{b}}{b_1}}{b_6}{b_8}
    }
    \infline
    \inference{
        \Gamma \vdash \transposeblock{z}{t'}{\epsilon}{b}{b_2}{b_3}
        &
        \Gamma \vdash \transposeblock{a}{t'''}{\cons{\cons{T}{\pi}}{t''}}{b_1}{b_4}{b_5}
        \\
        b_6 := \bindinfer{\pair{x}{\pair{z}{t''}}}{\paren{\accum{\ddot{x}}{y}{b_2}}}{
            \bindinfer{\pi}{\pair{x}{z}}{b_4}
        }
        \\
        b_7 := \bindinfer{\pair{x}{z}}{\pi}{
            \bindinfer{\pair{\dot{x}}{\dot{z}}}{\dot{\pi}}{b_3}
        }
        \\
        b_8 := \bindinfer{\pair{\pi}{\pair{t''}{t'''}}}{t}{
            \bindinfer{\pair{\dot{\pi}}{\unit}}{\paren{\accum{\ddot{\pi}}{\pi}{b_5}}}{b_7}
        }
    }{
        \Gamma \vdash \transposeblock{a}{t}{T}{
            \bindinfer{\pi}{\paren{\accum{\ddot{x}}{y}{b}}}{b_1}
        }{b_6}{b_8}
    }
\end{gather*}
\caption{Inference rules for transposition of other expressions and blocks.}
\label{fig:transpose-block}
\end{figure}

\end{document}